\documentclass[12pt, preprint]{aastex}

\begin{document}

\newcommand{\lya}{{\rm Ly}$\alpha$}
\newcommand{\Lya}{{\rm Ly}$\alpha$}
\newcommand{\lyb}{{\rm Ly}$\beta$}
\newcommand{\Lyb}{{\rm Ly}$\beta$}

\newcommand{\NHI}{\mbox{$N_{HI}$}}
\newcommand{\nhi}{\mbox{$N_{HI}$}}
\newcommand{\hi}{\mbox{\ion{H}{1}}}
\newcommand{\HI}{\mbox{\ion{H}{1}}}
\newcommand{\CI}{\mbox{\ion{C}{1}}}
\newcommand{\CII}{\mbox{\ion{C}{2}}}
\newcommand{\CIIstar}{\mbox{\ion{C}{2}$^*$}}
\newcommand{\CIII}{\mbox{\ion{C}{3}}}
\newcommand{\CIV}{\mbox{\ion{C}{4}}}
\newcommand{\MgII}{\mbox{\ion{Mg}{2}}}
\newcommand{\Mgii}{\mbox{\ion{Mg}{2}}}
\newcommand{\NI}{\mbox{\ion{N}{1}}}
\newcommand{\NII}{\mbox{\ion{N}{2}}}
\newcommand{\NV}{\mbox{\ion{N}{5}}}
\newcommand{\OI}{\mbox{\ion{O}{1}}}
\newcommand{\OIV}{\mbox{\ion{O}{4}}}
\newcommand{\OV}{\mbox{\ion{O}{5}}}
\newcommand{\OVI}{\mbox{\ion{O}{6}}}
\newcommand{\ovi}{\mbox{\ion{O}{6}}}
\newcommand{\SiII}{\mbox{\ion{Si}{2}}}
\newcommand{\SiIIstar}{\mbox{\ion{Si}{2}}$^*$}
\newcommand{\SiIII}{\mbox{\ion{Si}{3}}}
\newcommand{\SiIV}{\mbox{\ion{Si}{4}}}
\newcommand{\AlII}{\mbox{\ion{Al}{2}}}
\newcommand{\FeI}{\mbox{\ion{Fe}{1}}}
\newcommand{\FeII}{\mbox{\ion{Fe}{2}}}
\newcommand{\FeIII}{\mbox{\ion{Fe}{3}}}
\newcommand{\HST}{\mbox{\it HST}}
\newcommand{\ecs}{\mbox{$\rm erg~cm^{-2} sec^{-1}$}}

\newcommand{\simlt}{{\la}}
\newcommand{\simgt}{{\ga}}
\newcommand{\kms}{km~s$^{-1}$}
\newcommand{\cmsq}{cm$^{-2}$}

\slugcomment{accepted by {\it The Astrophysical Journal}}
\pagestyle{myheadings}

\title{The Low-Redshift Ly$\alpha$ Forest toward PKS~0405--123$^{1,2}$}

\author{Gerard M. Williger$^3$, Sara R. Heap}

\affil{Code 667, NASA Goddard Space Flight Center, Greenbelt MD 20771}

\author{Ray J. Weymann}
\affil{7610 San Marcos Ave., Atascadero CA 93422}

\author{Romeel Dav\'e}
\affil{Dept. of Astronomy, U. Arizona, Tucson AZ 85721}

\author{Erica Ellingson}
\affil{Center for Astrophysics \& Space Astronomy, 389 UCB, U. Colorado, Boulder CO 80309}

\author{Robert F. Carswell}
\affil{Institute of Astronomy, Madingley Road, Cambridge CB3 0HA, England}

\author{Todd M. Tripp}
\affil{Dept. of Astronomy, U. Massachusetts, Amherst MA 01003}

\author{Edward B. Jenkins}
\affil{Princeton U. Observatory, Princeton NJ 08544}


\begin{abstract}

We present results  for \Lya\ forest and metal absorbers from $\sim 7$~\kms\ resolution data
from the Space Telescope Imaging Spectrograph  for the QSO PKS~0405--123 ($z=0.574$,
$V=14.9$).  We analyze two samples of low redshift \Lya\ forest lines, a sample of
strong \Lya\ lines and a sample of weak ones.  The strong-line sample
consists of 60 \Lya\ absorbers detected at 4.0$\sigma$ significance with
column density $\log \NHI\geq 13.3$ over $0.002 < z < 0.423$; the
sample of weak lines contains 44 absorbers with a column density limit
of $\log \NHI\geq 13.1$ over $0.020 < z < 0.234$.
Seven of the \Lya\ absorbers show metal absorption lines.  Notably,
all of these metal systems appear to have associated \OVI\ absorption,
but the \OVI\ is often offset in velocity from the \Lya\ lines.  We do
not distinguish between metal and \Lya -only systems in the following
analysis, and use simulated spectra to aid in the interpretation of
results. The Doppler parameter distribution
for the strong sample has $\langle
b \rangle = 47 \pm 22$ km~s$^{-1}$.   For the weak sample, $\langle b \rangle = 44 \pm 21$
km~s$^{-1}$. Line blending and signal-to-noise effects likely inflate the Doppler parameters.
The redshift density is consistent with previous, lower resolution 
measurements for $\log \NHI\geq 14.0$. 
For absorbers with $13.1<\log \NHI< 14.0$, we find results consistent with previous high
resolution studies for $z<0.127$ but an overdensity of $\sim 0.2-0.3$~dex at $0.127<z<0.234$,
which we believe arises from cosmic variance. We find Ly$\alpha$-Ly$\alpha$
clustering in our sample on a scale of $\Delta v \leq 250$ \kms\ for $\log \NHI\geq 13.3$, 
which is consistent in strength and velocity scale with a numerical model of structure
evolution.
There is a void in the strong absorber sample at $0.0320<z<0.0814$ 
with probability of occurrence from a random redshift distribution
of $P<0.0004$. We detect line-of-sight velocity correlations of
up to 250 \kms\ between Ly$\alpha$ absorbers and 39 galaxies at $z<0.43$ in the field out to
transverse distances covering up to 1.6~$h^{-1}_{70}$ Mpc in the local frame.  The
Ly$\alpha$-galaxy two point correlation function is significant out to
$\Delta v < 250$~\kms\ and grows with minimum absorber \HI\ column
density, with the strongest signal for $\log \NHI\simgt 13.5-14.0$ absorbers.  The
strength is similar to that of the galaxy-galaxy correlation for galaxies of the same
mean luminosity as our sample, which implies that such \Lya\ absorbers have
masses $\log M/M_\odot = 11.3^{+1.0}_{-0.6}$.
The correlation  becomes insignificant for a sample limited to $13.1 \leq \log \NHI< 14.3$.
Including lower
column density systems  in the sample shows
correlations only out to $\Delta v < 125$~\kms , as would be expected for smaller
density perturbations.
We find a correlation between local galaxy counts
and local summed \HI\ column density, with peak significance on scales of $4000-6000$~\kms\
and the  probability of occurrence from uncorrelated data of $P=0.0009$ . 
Based on galaxy counts in the field, we predict regions of  low \HI\ column density at
$z\approx 0.45$ and high values at $z\approx 0.43$ and $z\approx 0.51$ toward PKS~0405--123.
Finally, we present column densities for a number of Galactic species.


\end{abstract}

\keywords{ cosmology: observations -- intergalactic medium -- quasars: absorption lines  }

\section{Introduction}

\footnotetext[1]{This paper is dedicated in memory of Ervin J. Williger, father of the first author,
who passed away on 2003 September 13.  His enthusiastic support and encouragement were essential to its successful
completion.}
\footnotetext[2]{Based on STIS IDT guaranteed time observations from the {\it Hubble Space Telescope}.}
\footnotetext[3]{present address:
Dept. of Physics \& Astronomy, Johns Hopkins U., Baltimore MD 21218}
An explanation for the \Lya\ forest is among the most impressive successes of CDM
structure formation models \citep[e.g.][]{Miralda96,Dave99,Dave01,DaveTripp01},  with
CDM simulations naturally reproducing the rich  \Lya\ absorption seen in QSO spectra. 
At $z<1.5$, where \Lya\ must be observed from space, a database has only slowly
accumulated for the \Lya\ forest, both for absorber statistics and  for comparisons
with the nearby galaxy distribution. 
The \Lya\ forest redshift density is known for higher equivalent
widths 
\citep[rest equivalent width $W_{0} > 0.200$ \AA ,][]{Weymann98}, but currently available samples are much smaller for weaker
\Lya\ absorbers with $W_{0} \ll 0.200$ \AA\ 
\citep[e.g.][]{Tripp98, Penton04, Richter04, Sembach04}.
 The \HI\ column density distributions at low and high
redshift
appear consistent with each other (\citeauthor{Penton04}),
whereas clustering in the \Lya\ forest at low $z$ is expected \citep{Dave99} but has
been challenging to detect. Among recent studies, \citet{Janknecht02} found no
evidence for clustering  via the two point correlation function for velocity scales
$\Delta v < 1000$~\kms\ for a sample of 235 \Lya\ forest lines over column densities
$13.1<\log \NHI < 16.4$ at redshifts $0.9<z<1.7$.  However, Penton et al. found a
$5.3\sigma$ signal for clustering at $\Delta v < 190$~\kms\ (and 7.2~$\sigma$ for
$\Delta v < 260$~\kms ) at $0.002<z<0.069$  for 187 \Lya\ absorbers with  rest
equivalent width $W_0>0.065$~\AA\  ($ \log \NHI \simgt 13.1 $).  The nature of
clustering in the \Lya\ forest is thus still a matter of active investigation, as is
the redshift density of low column density absorbers.

Numerical models for the \Lya\ forest have led to an interpretation for \Lya\ absorbers in
which weak lines are opacity fluctuations tracing the underlying dark matter distribution,
while stronger systems ($N(\HI )\simgt 10^{16-17}$cm$^{-2}$) represent 
higher overdensity regions and are often
affiliated with galaxies.  Thus \Lya\ absorbers
in a single quasar spectrum probe density perturbations across a wide range of scales, from
voids to groups of galaxies.   Metal absorbers provide complementary
information about the abundances, kinematics and ionization conditions near galaxy halos and in the
intergalactic medium \citep[e.g.][]{Bowen95,Dave98,Chen00,Tripp01,Tripp02,Tripp05,Rosenberg03,
Shull03,Tumlinson05,Keeney05,Jenkins05}.  Metal systems also
give indications about the strength
of feedback from supernovae in galaxies, and, in the case of \OVI\ absorbers, can contain large
reservoirs of baryons at both low and high redshift, e.g. \citet{Tripp00,Savage02,Carswell02,Simcoe04}
\citep[cf.][]{Pieri04}.  


Therefore, 
understanding the evolution of the intergalactic medium (IGM) near galaxies from high to low
redshifts will help constrain feedback processes that impact the formation of galaxies and 
their surrounding intragroup and intracluster media. 
At higher redshift,  \citet{Adelberger03} found that the IGM at $z\sim 3$  contains
more than the average amount of \HI\ at $1\simlt r\simlt 5\, h^{-1}$ comoving Mpc from LBGs.  
A strong
correlation between LBGs and metals (particularly CIV) in the IGM was also noted, indicating a
link between LBGs and CIV absorbers. They suggested that the decrease in \HI\ close to LBGs
and the metal enrichment around LBGs may be the result of supernovae-driven winds; the topic
currently remains an area of active investigation.

At $z \leq 0.5$, galaxies are relatively easy to identify, but the
\Lya\ forest itself is sparse.  In the nearby universe ($z < 0.3$),
many authors have found making a one-to-one association between
galaxies and \Lya\ absorbers to be a challenge.  Nevertheless, current
evidence indicates that \lya\ absorbers and galaxies cluster, though
less strongly than the galaxy-galaxy correlation \citep{Morris93,Stocke95,Tripp98,Impey99,Penton02}.
At even lower redshifts, \citet{Bowen02}  found 30 \Lya\
absorbers with the Space Telescope Imaging Spectrograph (STIS)  
around 8 galaxies, with column densities $\log N_{HI}\geq 13.0$;  sight
lines passing within $\sim 200\, h^{-1}$ kpc of a galaxy almost always show \hi\ absorption. 
The absorption tends to extend over 300-900 \kms , making a one-to-one correspondence between
an absorption system and any particular galaxy  difficult, though summing the absorption over
$\Delta v \sim 1000$~\kms\ produces a correlation of \hi\ strength with proximity to
 individual galaxies as found at $0.1\simlt z\simlt 1$ by \citet[][and references
therein]{Chen01}.  The correlation of \NHI\ with proximity to regions of
high galaxy density noted by \citeauthor{Bowen02} is likely related to the correlation
seen by \citeauthor{Adelberger03}  However, the quantitative 
evolution
of such correlations tracks overdensities associated with
typical \Lya\ absorbers of varying strengths, so that
at $z\simlt 1$, stronger lines are predicted to arise in regions of high galaxy
density  \citep[e.g.][]{Dave99,McDonald02}.

A more direct correlation arises between the  {\it sum} of the \hi\ column density over intervals of
$\sim 1000$~\kms\ and the {\it local volume density}\, of galaxies brighter than $M_B=-17.5$ \citep{Bowen02}
within $\sim 2\, h^{-1}$~Mpc of the QSO sight line, qualitatively similar to the Adelberger et
al. results over $5\,h^{-1}$ comoving Mpc.   The \hi\ column density -- galaxy volume density
correlation is intriguing, because the \Lya\ forest at $z\sim 3$ and $z<1$ appears to sample
different degrees of overdensities, very possibly due to the expansion of the universe and
rapidly falling UV background flux at
low redshifts \citep{Dave99}. A key to understanding the correlation would rely on  acquiring
high-quality observations of the \Lya\ forest -- galaxy correlation over  $0.1<z<1$ to bridge
the gap between the Bowen et al. and Adelberger et al. results.

In this paper, we address questions concerning the redshift density of low
redshift \Lya\ absorbers, especially low column density systems, their
correlations and their relationship to nearby galaxies by obtaining high
resolution UV spectra of the QSO PKS0405--12 ($V=14.9$, $z=0.574$) and
comparing them with a galaxy sample drawn from the literature and
from ground-based observations presented here.  It is one
of the brightest QSOs in the sky, and illuminates a much longer path in the
Ly$\alpha$ forest than most of the other QSOs of similar or brighter
magnitude.  PKS~0405--123 is therefore a prime target for studies of the low $z$
Ly$\alpha$ forest and metal absorbers and mid-latitude ($\ell=295^{\rm \, o} $,
$b=-42^{\rm \, o}$) studies of Galactic absorption.   
We use $\sim 7$~\kms\ resolution
STIS data to calculate the Ly$\alpha$ forest redshift density, Doppler
parameter, clustering and void statistics, and to cross-correlate the
Ly$\alpha$ forest and field galaxy redshifts.  
In support of the \Lya\ forest
work, we present observations of 42 galaxies
from a ground-based survey in the region, and add 31 galaxies from the 
literature to our sample.

Our observations and data reduction are described in \S~\ref{sec:observations},
and we outline the profile fitting method and absorber sample in \S~\ref{sec:absorbersample}.
The various absorber distribution functions and correlation functions 
are described in \S~\ref{sec:lyaforeststats}.  In \S~\ref{sec:lyagalcorrelations}, 
we compare the absorbers against a galaxy sample.
We discuss the results in \S~\ref{sec:discussion}, and
summarize our conclusions in \S~\ref{sec:conclusions}.
We adopt a cosmology of $H_0=70$ \kms\ Mpc$^{-1}$, $\Omega=0.3$, $\Lambda=0.7$ throughout this work.



\section{Observations and reductions}
\label{sec:observations}

Our principal data are from the {\it Hubble Space Telescope (HST)} plus STIS, using guaranteed time from Program 7576 
to the STIS Instrument Definition Team.  We also use archival \HST\ spectra of PKS~0405--123 
and ground-based galaxy images and redshifts in a supportive role.

\subsection{STIS spectra}

PKS 0405--123 was observed with \HST\ and STIS using grating
E140M for ten orbits (27208 sec)  on 1999 Jan 24 and 1999 Mar 7. We used  the $0.2''\times 0.06''$ slit
for maximal spectral purity.  For this setup, the STIS Instrument Handbook gives $R=45800$.

The data
were processed with 
{\sc calstis}\footnote{http://hires.gsfc.nasa.gov/stis/docs/calstis/calstis.html}, 
including correction for scattered light and spectral extraction, using STIS IDT
team software at Goddard Space Flight Center.  After merging the spectral orders, we
constructed a continuum using a combination of routines from 
automated {\sc autovp} \citep{Dave97} and interactive {\sc
line\_norm}\footnote{http://hires.gsfc.nasa.gov/stis/software/lib.html} 
by D. Lindler, because the regions around emission lines were
best done with manual fitting.  The final
spectrum has a signal-to-noise ratio per  $\sim 3.2$~\kms\ pixel in the \Lya\ forest of 4 to 7 except
for small dips at the ends of orders. Redward of 1581~\AA , we lose coverage of $\Delta z =
0.0035$ for \Lya\ due to six inter-order gaps.  The signal to noise ratio is $\sim 5-10$ 
per $\sim 2$ pixel
resolution element in the \Lya\ forest.

\subsection{HST archival spectra}

We retrieved and summed archival \HST\ spectra from the 
Faint Object Spectrograph (FOS, Proposal 1025, gratings G130H, G190H and
G270H)
Goddard High Resolution
Spectrograph (GHRS, Proposal 6712, gratings G160M and G200M) and
STIS (Proposal 7290, grating G230M), to provide consistency checks for wavelengths
covered by our E140M data and to provide complementary \Lya\ forest and
metal line information outside of the E140M spectral range.

\subsection{Ground-based galaxy images and spectra}

Direct images in Gunn $r$ and 
multi-object spectroscopic observations of galaxies in the PKS~0405--123 field
were made  on 1995 Jan 2 UT at the Canada-France Hawaii Telescope (CFHT), using
the MOS spectrograph, the O300 grism, and the z5 band-limiting filter.  The
data are part a survey of the environments of bright AGN \citet{Ellingson94}.
The stellar point spread function in the final stacked $10.2\times 9.8$~arcmin$^2$ MOS
direct image is
2.0~arcsec, and the limiting magnitude for extended sources is $r\approx 22.6$
($r=23.3$ for point sources). There are 175 galaxies with $r<22.5$, and 258
with $r<23.5$.
The spectral range was
5300--7000~\AA , with a resolution of about 12~\AA .  Object selection and
aperture plate design were carried out using the observation and reduction
techniques described in \citet{Yee96}. The band-limiting filter yielded 78
apertures on a single aperture mask spanning a region approximately 8 by 4
arcminutes around the quasar.  Slits were designed to be 1.5 arcseconds wide
and a minimum of 10 arcseconds long. Three one-hour integrations were taken
with this mask. PKS~0405--123 was targeted, preventing additional apertures
from being placed with about 5 arcseconds of the quasar.  The total area
covered in the data presented here 
is nearly three times that of \citeauthor{Ellingson94}, and the spectral
resolution is about double.

From these data, spectra of galaxies were identified via cross-correlation with
a series of galaxy templates,  and galaxy magnitudes in  Gunn $r$ were derived
from the MOS images used to create the aperture masks. 
Uncertainties in the photometry are about 0.1 mag.  Redshifts for objects up to
$r\approx 23$ were measured, with most of the fainter objects identified via emission
lines. The spectroscopic completeness is difficult to judge accurately because
some of the objects are stars and due to 
positional selection constraints from the
multi-slit spectroscopic setup. For $r<22.0$ we have spectroscopic redshifts
from our CFHT data for 46 of 206 objects (17\%), of which 126 are more extended
than our stellar point spread function. The targets were selected from an
8~arcmin square field centered on PKS~0405--123. The limited spectral range
provided somewhat higher success rates for objects with $0.35 < z < 0.7$, but
redshifts for galaxies between 0.1 and 0.35  were also identified (see
\citeauthor{Yee96} for  quantitative details). Also, because of the limited
wavelength coverage, many of the redshifts rely on a single line identification.
Uncertainties in galaxy redshifts are estimated to be about 100~\kms\ in the
rest frame, based on extensive redundant observations carried with MOS in this
and other surveys, with a possible systematic component less than about 
30~\kms .

To compare our results with those of \citet{Ellingson94}
our new galaxies for which we have redshifts have average $\langle r \rangle =
21.4\pm 0.9$ {\it vs.} $\langle r \rangle = 21.1\pm 0.8$ from
\citeauthor{Ellingson94} and \citet{Spinrad93} 
(excluding the uncertain magnitude of galaxy 309). 
The new CFHT data include 36 emission and
six absorption galaxies, with the faintest redshift for $r=23.4$, whereas the
sample from the literature includes 19 emission, ten absorption and two other 
galaxies with the faintest redshift for $r=22.5$. Three of the galaxies from the
literature, [EY94] 231 and 398, and [EY93] 309, were  re-observed
spectroscopically.  Results were consistent for the latter two objects. We
confirm a redshift of $z=0.5696$ for [EY94] 231, which coincides with [EY93]
232.  Overall, the galaxy list here contains redshifts for 44\% of the
87 galaxies in the MOS field to $r<21.5$ compared to 23\% for the smaller
area survey of \citeauthor{Ellingson94}, and
39\% of 175 galaxies to $r<22.5$ compared to 18\% for the previous work.






\section{Absorption line selection}
\label{sec:absorbersample}

The entire STIS E140M spectrum of PKS0405--123 is shown in Figure~\ref{fig:plotspec}.
We selected 
absorption features from the summed STIS E140M data at the $4.0
\sigma$ significance level with a Gaussian filter based on an {\sc autovp} routine, 
with half-widths of
8, 12, 16, and 20 pixels.  We then confirmed significant
features with a simple
equivalent width significance criterion based on a 4$\sigma$ threshold for contiguous 
pixels below the continuum, summed in quadrature.
Whenever possible and necessary, we confirmed spectral
features in the STIS E140M data with the archival
FOS and GHRS data.  
The STIS echelle spectra have substantially better spectral
resolution and wavelength coverage, so we did not sum together
data from the different instruments,
and only used the STIS E140M data for quantitative analysis of the \Lya\ forest.

G. Williger profile-fitted the data with {\sc vpfit}
\citep{Webb87} using Voigt profiles convolved with the STIS line spread
function (LSF) taken from the STIS Instrument Handbook.   Multiple transitions
(e.g. \Lya , $\beta$, $\gamma$) are used for simultaneous fits whenever
they improve constraints on column densities and Doppler parameters.
For comparison, R.
Carswell made an independent fit, though some of the higher column density
systems were not fitted as far down the Lyman series as done by Williger. The
two  profile fits were largely consistent.  We allowed for an offset
uncertainty in the continuum for some of the fits, as we deemed necessary for
accurate error estimates. In addition to \hi\ lines, we find and list,  for
completeness, profile fits or column density limits  \citep[from the apparent
optical depth method,][]{Savage91}  for a number of Galactic and  intervening
metal absorption lines. Our search for intervening metal lines  included
archival \HST\ GHRS G200M and STIS G230M data. The intervening metal absorber
redshifts will be considered with respect to galaxies in the region. However, a
further detailed study of them is beyond the scope of this paper; they are
discussed in \citet{Prochaska04}. 
Intervening systems with metals include $z =$ 0.09180, 0.09658, 0.167,
0.1829, 0.3608, 0.3616, 0.3633, and 0.4951.\footnote{We specifically searched
for \CIV\ and other metals from the high \HI\ column density complex at
$0.4057<z<0.4089$ in the GHRS and FOS data, but found nothing significant. 
However, there are weak, suggestive features for \CIV\ associated with the
complex in the FOS data.} We did not find significant \OVI\ absorption at
$z=0.1825$, in accord with the result of  \citet{Prochaska04}, and do not
include it in any of our analysis. \citet{Chen00} have studied the abundances
and ionization state of the $z=0.167$ system, and \citeauthor{Prochaska04} have
determined its \HI\ column density from FUSE data, which we adopt.

{\it Extragalactic metal systems.} \citep{Prochaska04} have
analyzed the metal absorbers in the PKS0405--123 spectrum in detail
including the systems at $z =$ 0.09180, 0.09658, 0.167, 0.1829,
0.3608, 0.3633, and 0.4951.  We find evidence for an additional
\ion{O}{6} doublet at $z = 0.3616$, with rest equivalent widths of
$46\pm 15$~m\AA\ and $44\pm 11$~m\AA\ for the $\lambda 1031$ and $\lambda 1037$
components, respectively, using the method of \citet{Sembach92}.  
It is interesting to note that
{\it all} of these metal systems show \ion{O}{6}.  However, in many
cases there are substantial offsets in velocity between the \ion{O}{6}
centroid and the \ion{H}{1} velocity centroid.  Based on our
Voigt-profile fitting centroids, we find that the \ion{O}{6} at $z =
0.16701$ is offset by $-34$ \kms\ from the corresponding \ion{H}{1}
centroid, \ion{O}{6} at $z=0.18292$ is offset by +51 \kms , and
\ion{O}{6} at $z = 0.36156$ is offset by +168 \kms .  Similar
differences in \ion{O}{6} and \ion{H}{1} velocity centroids have been
reported in other sight lines \citep[e.g.][]{Tripp00,Tripp01,Richter04,Sembach04}.
These velocity offsets are too
large to attribute to measurement errors.  Instead, they indicate that
the absorption systems are multiphase entities in which the \ion{O}{6}
and \ion{H}{1} absorption arises, at least in part, in different
places.

{\it Galactic absorption.} We find Galactic absorption from \HI , \CI , \CII , \CIIstar , \CIV ;
\NI , \OI , \AlII , \SiII , \SiIIstar , \SiIII , \SiIV , \ion{P}{2},
 \ion{S}{2}, \ion{S}{3}, \FeII , \ion{Ni}{2}.  
All Galactic absorption is at $-73 < v < 32$ \kms , so there is no
evidence for any high-velocity clouds. 
A more detailed analysis of these ions is
also beyond the scope of this paper. 
A complete list of absorber profile fitting parameters 
is in Table~\ref{tab:linelist}.  

\section{Simulated spectra to determine detection probabilities}
\label{sec:simulatedspectra}

To give us a clearer picture of our \Lya\ line detection probability as a 
function of s/n ratio and Doppler parameter,
we analyzed 1440 simulated
Ly$\alpha$ lines to determine the 80\% line detection probability. The simulated
lines were generated using the STIS LSF and a grid of values in Doppler
parameter $b$ ($17.5\pm 0.3$, $24.7\pm 0.3$, $35.0\pm 0.3$ \kms ) and in s/n ratio per pixel (4,7,12).   
The input
log~\nhi\  values range over the intervals 
12.85--13.35, 12.60--13.10 and 12.31--12.80 respective to the s/n ratio
grid.   The simulated spectra were continuum-normalized, so any errors, systematic or random,
involving continuum fitting would not be reflected in the subsequent analysis.
We calculated statistics comparing the line parameters 
for simulated lines which were successfully recovered {\it vs.} the input
values.  Results for the various combinations of input s/n ratio and Doppler parameter
are shown in Tables~\ref{tab:simulations_b} and \ref{tab:simulations_nhi}.
The mean recovered measured Doppler parameters match the mean input values within the
mean of the profile fitting errors.  The same is true for the log~\HI\ column
densities for input s/n ratio per pixel 12.  However, the mean recovered \HI\ column
densities are marginally high compared to the mean input values for s/n ratio 7 and 
are even less in agreement for s/n ratio 4, with a mean difference up to 
$\Delta \log \NHI = 0.17$ and
a mean measured column density error of $\langle \sigma(N(\HI )\rangle = 0.09$.
The reason for the marginally higher measured \hi\
column densities is because for the weak lines which 
we considered to probe the 80\%
detection threshold, noise effectively lowered an absorber's fitted column
density from above to below the detection threshold more often than it
increased an absorber's column density from below to above. Such selective
removal biases to higher values the \hi\ column densities for matches between simulated and
measured lines. The spurious detection rate is $<0.1$\% among all
the simulations above the 80\% detection probability threshold, and is
not considered significant. The boundary of the 80\% detection probability threshold 
can be parametrized
by $\log \NHI = 12.870 +0.344 \log (b/24.7) - 1.012 \log ({\rm
snr}/7.)$, where snr is the s/n ratio per pixel.

\section{Ly$\alpha$ absorber statistics}
\label{sec:lyaforeststats}

In consideration of the \nhi\ sensitivity of the data and to provide a comparison
to other studies in the literature,
we define two
samples based on \hi\ column density:
a ``strong" one for $\log (N(\HI ))\geq
13.3$ and a ``weak" one for $\log (N(\HI ))\geq
13.1$.  The strong sample was chosen to take advantage of the minimum s/n ratio over most of
the \Lya\ forest region, while the weak one was chosen to focus on weak absorbers and to be
comparable to other high resolution studies.  We use the s/n ratio of the data as a function of wavelength, 
a $4.0\sigma$ significance
threshold and a Doppler parameter
of $b=40$~\kms , which is close to the mean value of $b$ in our data\footnote{
Note that by choosing $b=40$~\kms , this leads to a higher, more conservative 80\%
probability detection threshold in \NHI\ by 0.07 dex than by choosing $b=25$~\kms , which is more typical
of high resolution, high s/n ratio \Lya\ forest data.  We prefer to be conservative given
the possibility of unresolved blends due to our s/n ratio, discussed in \S~\ref{subsec:dopplerparams}.
This choice of $b$ does not affect our subsequent analysis in any way except for this small change
in threshold $\log \NHI$.} to determine the redshifts corresponding to the two samples.
Consequently, the strong sample contains the range $0.002<z<0.423$ and 60 \Lya\ absorbers, whereas
the weak one spans $0.020<z<0.234$ and includes 44 absorbers.
We show the Doppler parameters and \HI\ column densities for all \Lya\ forest systems
we detect plus the detection sensitivities
at the 50, 60 and 80\% levels as parametrized above for the strong and weak
samples in Figure~\ref{fig:nb}.
We checked for misidentifications of our \Lya\ forest sample against the metal line list of
\citeauthor{Prochaska04}, and found only five instances of potential overlap, at
$z_{Ly\alpha }=$ 0.107973, 0.173950, 0.205024, 0.205407, 0.386720.  None of the \Lya\ absorbers falls in
either the strong or weak samples, and all of the features listed in \citeauthor{Prochaska04} are
listed as upper limits rather than secure metal line detections.

We examine the Ly$\alpha$ forest Doppler
parameter distribution, redshift density $dN/dz$, void distribution, and clustering via the two point correlation
function for the strong and weak samples.
We do not distinguish between
metal and Ly$\alpha$-only systems in this analysis, in part because
metals are found in progressively lower column density \Lya\ systems
\citep[e.g.][and references therein]{Cowie95, Songaila96, Pettini01}, 
the absorbers we can detect at low redshift probably are best
related to
large density perturbations at $z\simgt 2$ \citep{Dave99} which are commonly associated
with metals, and because the limited wavelength range of the 
data does not permit uniform coverage for metal
detection.

\subsection{Doppler parameter distribution}
\label{subsec:dopplerparams}

We parametrized the \HI\ column density distribution as $dN/dN(\HI )\propto \NHI
^{-\beta}$. For the weak sample, $\beta = 1.96\pm0.15$ with a probability that the
data are consistent with a power law $P=0.08$.  For the strong sample, 
$\beta=2.06\pm
0.14$ and $P=0.023$.

The Doppler parameter distribution for the strong 
sample has mean, median and standard deviation of
47, 47, 22~\kms , respectively.
The Doppler parameter distribution for the weak 
sample has mean, median and standard deviation of
44, 44, 21~\kms .
The large mean arises from a tail at large
values, which likely results from unresolved blends, and exists in both of the
profile fit sets from Carswell and Williger (Fig.~\ref{fig:compare_rfc_gmw}). There is a 
weak trend for the
higher Doppler parameter lines above the 80\% detection thresholds
to have lower \HI\ column densities (Fig.~\ref{fig:dndb}), which 
runs counter to what is observed in higher s/n ratio data and theoretical
expectations \citep{DaveTripp01}.  A possible cause is
blending of weak lines which tend not to be resolved for
signal-to-noise ratios near the $\log \NHI>13.3$ 80\% probability detection threshold.
If we divide the strong sample into high and low \HI\ column density halves,
which occurs at 
$\log \NHI= 13.485$ for the strong sample,
and perform a Kolmogorov-Smirnov test to determine the likelihood that the Doppler parameters
are drawn from the same distribution, the probability is only $P=0.003$.  
We only find two systems with $b>85$~\kms ,
both in metal absorption complexes in which both \Lya\ and \Lyb\ profiles were
fitted simultaneously:
$z=0.166962$, $\log \NHI = 14.17\pm 0.06$, $b=112 \pm 8$~\kms\ and
$z=0.361025$, $\log \NHI = 14.35\pm 0.04$, $b=134 \pm 17$~\kms .
If we exclude
the two $b>85$~\kms\ absorbers, which we suspect are likely to be unresolved blends or
are poorly separated from $\log \NHI> 15$ systems, the probability decreases 
to $P=0.002$.
However, the weak sample, which divides in half at $\log \NHI= 13.34$, 
shows no such indication for Doppler parameter distribution 
differences between the
low and high column density subsamples 
($P=0.86$ and 0.72, depending on whether 
the two $b>85$~\kms\ systems are included or excluded).
The strong sample is drawn from spectral regions where the signal-to-noise ratio is
lower on average than from the weak sample, 
which supports the line blending explanation for high Doppler parameters.
It is noteworthy how blending remains a problem for absorption line work even at
low redshift, despite the apparent sparseness of the \Lya\ forest.

\subsection{Redshift density}
\label{subsec:redshiftdensity}

The Ly$\alpha$ forest redshift density has been studied intensively for over 20
years.  We consider strong and weak \Lya\ absorbers in turn,  and compare our
data against  recent \HST\ observations \citep{Penton04}, VLT UVES observations
\citep{Kim02} and other ground-based and \HST\ values  of varying resolution
from the  literature.  We made
several corrections to the redshift path.  First, inter-order gaps
block $\Delta z = 0.0048$ from the strong survey (though none for the weak, because the
lowest wavelength gap is outside the weak survey range at $z_{Ly\alpha}=0.315$).
We then address the decrease in \Lya\ absorber sensitivity for Galactic, intervening
metal and higher order Lyman lines.  We deem metal and higher order Lyman
lines which produce $\tau\geq 2$ as blocking reliable detection of \Lya\ lines, based
on our experience with profile fits.
Galactic and intervening metal lines would then affect
$\Delta z = 0.0024$ and 0.0018 for the strong and weak surveys, respectively.
\Lyb\ and higher order lines from known systems produce $\tau \geq 2$ for
$\Delta z = 0.0005$ for both the strong and weak surveys.  
If we were to choose $\tau\geq 1$ as our blocking threshold, the amount of blocked redshift
space would increase by 50\% for
the metals and double for the higher order Lyman lines, which is still an effect only
on the order of 1\% .
The net result is
that the unblocked redshift paths for the strong and weak samples are
$\Delta z = 0.413$ and $\Delta z = 0.212$, respectively.


\subsubsection{High column density lines}

We find 12 strong ($\log \NHI \geq 14.0$)  
\Lya\ absorbers at $0.002<z<0.423$  ($\log dN/dz
= 1.46^{+0.11}_{-0.14})$.  
The 
redshift density is consistent within the errors with both the low resolution survey carried
out by  \citet{Weymann98}  and the medium resolution surveys of \citet{Penton04} and
\citet{Impey99} (Figure~\ref{fig:plotkim}, right panel).


\subsubsection{Low column density lines}

For absorbers with $13.1\leq \log \NHI < 14.0$,
we find 38 systems over $0.020<z<0.234$ ($\log dN/dz = 2.25^{+0.07}_{-0.07}$). 
Given the large number of absorbers, we split our low column density sample's
redshift range in half at $z=0.127$, and detect 13 and 25 lines in the lower and
upper halves ($\log dN/dz = 2.08^{+0.11}_{-0.14}$ and  $\log dN/dz =
2.37^{+0.08}_{-0.09}$ respectively).  Our lower half redshift sample is then
consistent within $1\sigma$ errors with the  samples of \citet{Sembach04} and
\citet{Richter04} who used the same instrumental setup\footnote{Data for a high
resolution sample toward 3C~273 will be discussed in Williger et al. (2005, in
prep.).} as we did, and  consistent with the \cite{Penton04} result at the $2\sigma$
level (Figure~\ref{fig:plotkim}, left panel).   The higher redshift half of our data
is consistent with the redshift density of \citet{Richter04} at $0.12\simlt z \simlt
0.24$ at the $1.4\sigma$ level, but is higher by a factor of 3.4 compared to the
result from \citeauthor{Penton04} based on 15 sight lines of $\log dN/dz|_{z=0} =
1.85\pm 0.06$, which are of lower resolution ($\sim 41$~\kms\ for STIS+G140M, $\sim
19$~\kms\ for GHRS+G160M). The lower $dN/dz$ from the \citeauthor{Penton04} results
may arise from uncertainties in their  line detection sensitivity. Figure 4 in
\citeauthor{Penton04} shows a very steep increase in the rest equivalent width
distribution at $\log \NHI< 13.4$. A small uncertainty in the line detection
detection  threshold of $\sim 0.1$~dex at such low column densities could easily
produce an uncertainty in the redshift density on the order of a factor of 2-3. A
complicating factor may be their lower resolution and assumed a constant Doppler
parameter of $b=25$~\kms , which could have biased the column densities to be low
for their weak absorbers. There are also blending effects to consider with the G140M
data.   \citeauthor{Penton04} made an in-depth discussion of the effects of
comparing $dN/dz$ data of differing resolutions, and showed that  low resolution
{\it HST} FOS Key Project data of resolution $\sim 240$~\kms\ overestimated the
number of high equivalent width absorbers.  A similar effect may underestimate low
equivalent width (or column density) absorbers in the data of \citeauthor{Penton04}
compared to ours.

Within the PKS~0405--123 data, the $dN/dz$ difference between the lower and higher
redshift samples reflects the large proportion of lines (13 of 25) at
$0.127<z<0.175$ (Fig.~\ref{fig:plotdndz_weak}). The redshift interval contains
groups of five absorbers at $0.1309<z<0.1365$,  three at $0.1513<z<0.1530$ and three
at $0.1612<z<0.1631$.  Most of the \HI\ column densities  (10/13) at $0.127<z<0.175$
are in the range $13.3\leq \log \NHI\leq 14.0$, which corresponds to $\delta \rho /
\rho \approx 3-10$ \citep{Dave01}.    We note that the difference in $dN/dz$ within
the PKS~0405--123 sight line is significantly greater than that for PG~1259+593
(\citeauthor{Richter04}) as well as for the HE~0515-4414 sight line ($0.99<z<1.68$,
\citeauthor{Janknecht02}).  The PKS~0405--123 sight line could  be an extreme example
of cosmic variance, which will be considered in more detail in the discussion
(\S\ref{sec:discussion}).

\subsection{Broad lines}
\label{subsec:broadlines}

There have been detections of broad \Lya\ absorbers with $b>40$~\kms\ which may
sample intervening warm-hot IGM (WHIM) gas \citep[e.g.][and references
therein]{Bowen02,Penton04,Richter04,Sembach04}.  However, a number of factors can
yield broad lines, including blending, low s/n ratio, kinematic flows, Hubble
broadening and continuum undulations.  Following the example of
\citeauthor{Richter04} and \citeauthor{Sembach04}, we count 34 broad \Lya\ lines
with $40<b<100$~\kms\ in our strong sample and 27 in our weak one, making
$dN/dz=82\pm 14$ and $127\pm 25$ respectively. \citeauthor{Sembach04} find
$dN/dz\sim 40\pm 16$ for rest equivalent width $W_r>30$~m\AA\ ($\log \NHI\simgt
12.6$ for $b=20$~\kms ), and \citeauthor{Richter04} find $dN/dz\sim 23\pm 8$ for
$W_r>45$~m\AA\ ($\log \NHI\simgt 12.9$ for $b=20$~\kms ).  The large redshift
density for broad lines in our sample would be consistent with the effects of  a low
s/n ratio and line blending.

\subsection{Clustering and voids}
\label{subsec:clusteringandvoids}


\subsubsection{Clustering}

Clustering in the Ly$\alpha$ forest has been studied in a number of
cases \citep[][and references therein]{Janknecht02}, with  
weak clustering indicated on velocity scales of $\Delta v < 500$ \kms .  
We use the two point correlation function $\xi(\Delta v)\equiv
[n_{obs}(\Delta v)/n_{exp}(\Delta v)] - 1$ where $n_{obs}$ and
$n_{exp}$ denote the observed and expected numbers of systems in a
relative velocity interval $\Delta v$.  

We created 10$^4$ absorption line lists using Monte Carlo
simulations weighted in $z$ using the slow redshift evolution
derived by \citet{Weymann98}, $dN/dz \propto (1+z)^\gamma$,
$\gamma = 0.26$.
For each simulation we drew a
number of absorbers from a Poissonian distribution with a mean equal
to the number actually observed in our sample, and tested for
correlations for a grid of \HI\ column densities, moving up from
$\log \NHI \geq 13.1$ in increments of $\Delta \log \NHI = 0.1$, using the appropriate
redshift range sensitivity depending on the minimum column density.  The strongest signal
comes for the strong sample ($\log \NHI \geq 13.3$), 
in which 15 pairs are observed with
velocity separation
$\Delta v\leq 250$~\kms , and $7.7\pm 2.9$ are expected.  

All of the pairs with $\Delta v<250$~\kms\ were 
inspected for noise spikes, potential metal line contamination 
and other possible sources of misidentification.
They are listed in Table~\ref{tab:lyapairs} and plotted in Fig.~\ref{fig:plot_lyapairs}.
We note that six of the pairs lie in
the same redshift range which produces the high redshift density at $0.13\simlt z \simlt 0.17$
described in \S~\ref{subsec:redshiftdensity}.  There are eight true pairs and three
sets of triplets.
Only the triplet at $z\approx 0.167$ is known to have metals.  It and the
triplet at $z\approx 0.351$ contain absorbers with
$b>60$~\kms .  We show below that such high Doppler parameter lines 
may be produced by unresolved blends.

The probability of unclustered data producing a signal of the strength we see  in the
$\Delta v < 250$~\kms\ bin is $P=0.014$, or $2.5\sigma $ significance.  The signal
decreases at lower and higher $N(\HI )$ threshold, because of decreased numbers
on the high side, and on the low side a combination of 
decreased close pair
detection efficiency as we reach our 80\% detection threshold and 
weaker expected
clustering among lower density perturbation \Lya\ systems. We thus only have a lower
bound to the correlation significance, because our control sets have not been
filtered to reflect the true ability to resolve close absorption lines in our data,
which we explain below.

The closest pair of absorbers we observe anywhere in 
our data has a velocity difference of 44~\kms\
($z=0.029852, 0.030002$, $\log \NHI = 13.21, 14.18$). The minimum splitting in
the strong survey is 45~\kms , for $z=0.166962, 0.167139$, which is just as small
within the redshift errors. In principle, we may be
able to resolve closer pairs, given our resolution of $\sim 7$~\kms , but the exact limit is
a complex function of signal-to-noise  ratio, absorption line parameters, availability of
\Lyb\ and higher order transitions, presence of Galactic or intervening metal lines etc. To
probe the resolution sensitivity in a rough fashion,  we produced random spectra as in our
simulations above with s/n ratio 5, which corresponds to the minimum signal-to-noise ratio
in our weak line survey, with no pairs permitted for velocity splittings $\Delta v \leq
15$~\kms (roughly twice the spectral resolution). From input sets of \Lya\ lines with
$13.2\leq \log \NHI\leq 13.3$ and $40<b<41$~\kms ,  which should be above our 80\%
detection threshold based on our previous simulations, we found 34 pairs with input values
of $15<\Delta v<100$~\kms .   The smallest velocity splitting for which both components were
successfully profile-fitted in the simulations 
with $\log \NHI\geq 13.2$ was 37~\kms , which is very close
to our minimum observed value.  For $\Delta v \leq 37.0$~\kms , 0/13 pairs had two
components detected for any column density at all.
For $37.0<\Delta v \leq 60.5$~\kms , 
4/9 pairs showed two components, but each of the pairs had one component fitted to $\log
N(\HI )<13.2$ due to noise effects.  Such pairs would be completely recovered with a higher
local s/n ratio, higher \HI\ column density or lower Doppler parameter. For $60.5<\Delta v \leq
100$~\kms , 2/12 pairs only showed one component with a profile fit, 4/12 showed two
components but one with  $\log \NHI<13.2$, and 6/12 pairs had both components
successfully recovered above our column density threshold. Line parameters were recovered
well when both components of a pair were detected: column density mean
$\langle \log \NHI \rangle 
= 13.41\pm 0.13$,
and
range $13.22\leq \log \NHI \leq 13.61$, and Doppler parameter mean
$\langle b \rangle = 40\pm 12$~\kms\ 
and 
range $23\leq b \leq 61$~\kms .  In the cases of completely blended lines, however, the
column densities and Doppler parameters were understandably inflated: $\langle \log \NHI
\rangle
= 13.62\pm 0.11$, with a
range $13.20\leq \log \NHI \leq 13.72$, and 
$\langle b \rangle = 51\pm
18$~\kms , with a range $35\leq b \leq 111$~\kms .  
The combined column density of a blend is
consistent with the sum of the individual column densities \citep{Jenkins86}. 

Unresolved
\Lya\ pairs can therefore account for at least a fraction of the two $b>100$~\kms\ absorbers
in our data (\S~\ref{subsec:dopplerparams}). Additional simulations were made to verify that
the pair detection sensitivity predictably decreases for even lower column density input
lines. Based on the results of our simulated spectra,
we conclude that the pair resolution sensitivity at our 80\% detection threshold 
increases from zero at $\Delta v\approx 35$~\kms\ to roughly 50\% by $\Delta v \approx
100$~\kms , and at higher velocity separations presumably increases to $1-P^2$ where $P$ is
the local detection probability when the line separation is comparable to the line widths.
The resolution of closer pairs toward other sight lines is likely the result of higher s/n ratio
in other data
e.g. 7--17 per resolution element for PG~1259+593 \citep{Richter04} and 15 per resolution
element for PG~1116+215 \citep{Sembach04}.


We therefore filtered our simulated line lists so that one member of a pair with $\Delta
v<43$~\kms\ was eliminated, which affected 1.8\% of the input lists, which is $<1$ line per
simulated line list.   The number of pairs per velocity bin per
simulated line list is renormalized to the number of observed
pairs in any case, so there should be no adverse effect on the
statistics, except for eliminating the number of expected pairs at
$\Delta v<43$~\kms . We then find a two point correlation function
value  for $\log \NHI\geq 13.3$  at $\Delta v<250$ \kms\ of
$\xi(\Delta v) = 1.2$, with 15 pairs observed and $6.8\pm 2.7$
expected (Fig.~\ref{fig:twoptcorr}), which is an overdensity
of $3.1\sigma$ and
has probability
$P=0.005$ to be matched or exceeded by the Monte
Carlo simulations.   This is still a lower bound, because our
detection efficiency is $<100$\% for $43<\Delta v \simlt 100$~\kms .  
Further refinements to the pair detection sensitivity would
best be done with a large set of simulated spectra to correct for
the overestimated number of pairs  at $40\simlt \Delta v \simlt
100$~\kms\ in our Monte Carlo simulations,  which should involve
using \Lyb\ and higher order lines where possible, but the
correction should be on the same order or less than the simple
correction above, because the fraction of undetected pairs would
be $<100$\% .

\citet{Dave03} performed a hydrodynamic $\Lambda$CDM simulation with box size 22.222~Mpc~$h^{-1}$
at $z=0$
to predict the effects on
\Lya\ forest correlations of
bias in the relationship between \NHI\ and the underlying dark matter density.
Our two point correlation results agree with the simulation results
within $1.2\sigma$ for our signal at $\Delta v<250$~\kms\
(Fig.~\ref{fig:twoptcorr}).  The agreement
reinforces the bias evolution predictions in those models, despite 
the complex physics in the evolution of the IGM at $z<1$.  


\subsubsection{Voids}

Voids in the \Lya\ forest are rare, and
have mainly been studied at higher redshift.
We searched for regions in velocity space
devoid of Ly$\alpha$ systems using the same
10$^4$ Monte Carlo simulations, and find evidence for a void at
$0.0320<z<0.0814$ ($\Delta v = 14000$ \kms , 206$h^{-1}$ comoving  Mpc) for $\log {\rm
N_{HI}}\geq 13.3$.  The probability for such a void to be matched or exceeded in
velocity space is $P=0.0004$, based on a sample of 
over $5.7\times 10^5$ \Lya\ absorber spacings.  For ${\rm
N_{HI}}\geq 13.2$, we find two voids of similar significance at
$0.0320<z<0.0590$ and $0.1030<z<0.1310$ ($\Delta v = 7744$, 7512 \kms , 
$P=0.007$), while for $N(\HI )\geq 13.1$, the lower redshift
void persists ($0.0320<z<0.0590$, $P=0.003$).  We will compare low and high redshift
voids in \S~\ref{sec:discussion-clustersvoids}.

\section{Ly$\alpha$-galaxy correlations}
\label{sec:lyagalcorrelations}

\citet{Spinrad93} and \citet{Ellingson94}
surveyed for galaxies in a
$10\times 8$ arcmin$^2$ field around PKS~0405--123.  We use
18 of their published redshifts at $z<0.47$, which is $10^4$~\kms\ redward of 
the long wavelength end of our STIS data,
(deferring to the later
paper in case of redshift disagreement)
plus nine more out to the QSO redshift of $z=0.58$ and four background galaxies to $z=0.66$.  
The CFHT data reveal a number of additional
galaxies in the 
field, 22 of which are
at $z<0.47$.  We have a sample of 40 galaxies at
$z<0.47$ (plus 19 more out to the PKS~0405--123 host cluster at $z=0.58$) 
covering $r$-magnitudes $18.6<r<22.9$ ($\langle r \rangle =
20.9\pm 1.0$) and absolute magnitudes  of $-22.6<M_r<-13.0$ 
($\langle M_r \rangle = -19.8\pm 1.7$, using the distance
modulus for our adopted cosmology).  The sample at $z<0.47$ is dominated by
32 emission line galaxies, along with six absorption line galaxies
and two of unknown spectral type.
Galaxy positions, redshifts, magnitudes, types, and impact parameters 
for both new presentations and galaxies from the literature
are listed in
Table~\ref{tab:galaxies}, with a direct image 
in Figure~\ref{fig:galaxies}.  Plots of the galaxies in relation to the absorbers 
in redshift and right ascension/declination
are in Figures~\ref{fig:galconera} and \ref{fig:galconedec}.


\subsection{Two point correlation function}

We cross-correlated our galaxy sample with the Ly$\alpha$ forest, 
including all galaxies within 1.6$h^{-1}_{70}$ Mpc in
the local frame for $z<0.47$.  For the strong \Lya\ forest sample,
there is a signal in the two point correlation function
at $\Delta v \leq 250$ \kms , ($\xi(\Delta v)=2.2$, 
$5.8\sigma$ significance, 33 \Lya -galaxy pairs observed, $10.2\pm 4.0$
expected).  The significance of the absorber-galaxy correlation
peaks for $\log \NHI\geq 13.5$ (Fig.~\ref{fig:lyagalcorr}), and
the value of the correlation itself is maximum at  $\log \NHI\simgt 13.9-14.0$ 
(Fig.~\ref{fig:lyagalcorr_grid}).
The correlation at $13.5\simlt \log \NHI \simlt 14.0$ 
is nearly that of the galaxy-galaxy correlation function for
our sample \citep[$\langle M_r\rangle = -20$,][]{Zehavi04}.
The significance of the correlation
declines at the high end due to progressively
smaller sample sizes of \Lya\ absorbers, though
the actual strength of the correlation tends to stay high.
For minimum column density thresholds $13.1 < \log \NHI< 13.2$, in which we only
consider the high s/n data at $0.020<z<0.234$, 
the strength decreases, 
presumably because lower mass systems are included in the calculation.
For absorbers with $\log \NHI\geq 13.1$
the correlation
maximum is only significant for $\Delta v < 125$~\kms , with a signal at the $4.4\sigma$
level (9 pairs observed, $2.2\pm 1.6$ expected).  
Such behavior would be consistent with higher column density systems having longer correlation
lengths with galaxies, which would be expected from larger density perturbations.
If we limit the upper column density threshold, the significance
drops from $>3\sigma$ to $2.7\sigma$ for $13.1 < \log \NHI< 14.3$, though that is likely in part due
to the small sample size (6 pairs observed, $2.0\pm 1.5$ expected).  Nevertheless, the implication is
that
the correlation comes from relatively high column density ($\log \NHI\simgt 13.5$) absorbers,
corresponding to $\log \rho/\bar{\rho} \simgt 0.9$ at $z=0.2$ \citep{Dave99}.

The galaxy distribution with redshift yields clues to understanding
the observed correlation. There is a galaxy overdensity around the
$z=0.1671$ partial Lyman limit system, with another around the
metal absorbers at $z\approx 0.36$, both of  which both contribute
significantly to the number of galaxy-absorber pairs. Conversely,  
there is only one galaxy (at $z=0.0791$) in any of our listed
Ly$\alpha$ forest voids within our CFHT field.  We performed a
similar correlation test for the six \OVI\ absorbers at $z<0.43$
\citep[including those listed in][]{Prochaska04},
noting that the systems at $z=$ 0.09180, 0.09658, 0.16701 and
0.36335 are within $\Delta v<250$~\kms\ of galaxies, and that those
at $z=$ 0.18292 and 0.36156 are not. For $\Delta v < 250$~\kms , we
observe 6 absorber-galaxy pairs and expect $1.1\pm 1.4$, with  a
$P=0.016$ chance of having an equal or greater number of
absorber-galaxy pairs in that bin. In spite of the paucity of our
\OVI\ sample,  our results are consistent with those of
\citet{Sembach04} and Shull et al. (2005, in prep.) 
in that \OVI\ absorbers are  not randomly
distributed with respect to galaxies.  We note that of the five
galaxies within 250~\kms\ of \OVI\ absorbers (numbers 22, 33, 45,
54 and 60), at least one of them is very luminous ($M=-22.3$), which
could signal a particularly deep potential well, and the two at
$z\approx 0.167$ are within 118~$h^{-1}$~kpc of the sight line to
PKS~0405--123, by far the closest of any galaxies within $\Delta v
\leq 250$~\kms\ of a strong \Lya\ absorber.


\subsection{Correlation with galaxy density}

\citet{Bowen02} 
suggested a correlation between the density of
Ly$\alpha$ components along a sight line and the volume density of
$M_B<-17.5$ galaxies within $\sim 2$ Mpc.   Despite the
incompleteness of our galaxy counts, and the STIS \Lya\ forest
data being limited to $z<0.42$, there is a striking correlation
between the local galaxy density and local \HI\ column density in
the \Lya\ forest  (Fig.~\ref{fig:nhigalhist}). As a very basic
test of whether the rank of high \HI\ column densities correlates
with high local galaxy counts in a series of redshift bins, we
performed the Spearman and Kendall rank tests for redshift bins
of  $0.010 \leq \Delta z \leq 0.035$ in increments of  $\Delta z =
0.005$. The mean Spearman rank coefficient is $\langle \rho\rangle
= 0.59\pm 0.09$,  with a mean two-sided significance level of its
deviation from zero of $\langle p_\rho\rangle =0.010$.   As
a check, the  Kendall rank coefficient is similar, with a mean 
$\langle \tau \rangle = 0.47 \pm 0.08$,  with mean two-sided
significance $\langle p_\tau \rangle =0.007$. 
It is therefore
unlikely that \NHI\ and local galaxy counts are uncorrelated.
The maximum signal occurs for bins of $\Delta z =
0.02$ ($4000<\Delta v < 6000$~\kms\ over $0<z<0.4$), in which the
Spearman significance is $p_\rho=0.0009$ and Kendall significance
is $p_\tau =0.0004$.

We tested whether  the difference in volume surveyed for galaxies
may play a role in skewing the rank correlation statistics,
because our volume sample at low redshift is smaller than that at
high redshift. The total volume surveyed in a $5\times
5$~arcmin$^2$ field over $0<z<0.42$ is
$3100h^{-3}$~comoving~Mpc$^3$, of which 98.6\% is at $z>0.10$ and
88.5\% is at $z>0.20$. If the data for $z<0.10$ are ignored, the
Spearman rank coefficient is  $\rho = 0.65$ with $p_\rho=0.007$,
with the corresponding Kendall rank correlation  $\tau = 0.55$,
$p_\tau = 0.003$.   If data for $z<0.20$ are excluded, $\rho =
0.88$ with $p_\rho=0.0003$  and $\tau = 0.77$, $p_\tau = 0.001$. 
Removing data at $z>0.20$ from the sample leaves $\rho = 0.77$
with $p_\rho=0.016$  and $\tau = 0.65$, $p_\tau = 0.015$. The
correlation therefore appears consistent over a range of redshift
subsamples. We conclude that the difference in galaxy survey
volume as a function of redshift  makes no significant difference
to the correlation between \HI\ column density and galaxy number
counts.


\section{Discussion}
\label{sec:discussion}

\subsection{Doppler parameter and redshift distributions}

Our Doppler parameter distribution is likely affected by
unresolved blends.  However, the total column densities in such
complexes should not be far in error \citep{Jenkins86}, which we
also conclude from our small set of simulated close \Lya\ lines.
We  do not find clear evidence for broad ($b>100$~\kms ) \Lya\
absorbers which are not unresolved blends or results of poor
continuum fitting, either of which could be exacerbated by the
limited s/n ratio.

The redshift density toward PKS~0405--123 at $\log \NHI > 14.0$ 
is consistent with the low resolution \HST\ Key Project data of
\citet{Weymann98} and with the medium resolution \HST\ G140L data
of \citet{Penton04} amd \citet{Impey99},  both of which employ
larger \Lya\ system samples than ours.  \citeauthor{Penton04}
discussed the effect of resolution on $dN/dz$ in detail,
concluding that lower resolution data, such as from the FOS
(including the Key Project),  somewhat increases the number of
$\log \NHI>14.0$ absorbers.  Our data, which are of higher
resolution than either the G140M or FOS samples, appear to bear
this out, though the difference is within the $1\sigma $ errors,
as is the predicted value of the \Lya\ forest redshift density at
$z=0$ from $\Lambda$CDM simulations by \citet{Dave99}.

The $z<0.127$ half of the $13.1<\log \NHI<14.0$  sample  is
consistent both with  the redshift densities of the STIS E140M
samples of \citet{Sembach04} and \citet{Richter04} at $z\simlt
0.12$, and the $1.0\simlt z \simlt 1.7$ data of
\citet{Janknecht02}, despite containing one to two voids,
depending on how they are defined.  The $z<0.127$ redshift
density toward PKS~0405--123 is also within $1.6\sigma$ of the result
of \citeauthor{Penton04}, which covers over five times the
redshift path that our sample does, but may suffer from blending
problems.  

For the $0.127<z<0.234$ low column density $dN/dz$, cosmic
variance is a possible explanation for both the anomalously high
redshift density and the \Lya -\Lya\ and \Lya - galaxy clustering
strengths. The redshift density difference between the two halves
of our sample ($\Delta \log dN/dz = 0.34$) is slightly larger than
the difference in the sample of  \citeauthor{Penton04}  ($\Delta
\log dN/dz = 0.27$), who divided their data into eight bins over
$0.002<z<0.069$ (their Fig.~7).  The sample of \citet{Richter04}
also exhibits a smaller but still pronounced variation of $\Delta
\log dN/dz = 0.13$ between $z\simlt 0.13$ and $0.13\simlt z \simlt
0.25$. If we subdivide our redshift bins into intervals of $\Delta
z \sim 0.04$, the variation is  $\Delta \log dN/dz \approx 0.6$
(Fig.~\ref{fig:plotdndz_weak}). The large $\Delta \log dN/dz$
variations in the PKS~0405--123 data lead us to believe that this
may be a typical level of cosmic variance  for the sight line. 

Unresolved blends could also play a role in our high $dN/dz$
result, if they transfer more low column density lines to $\log
N(\HI )<13.1$ (``blending out") than produce  high  lines at $\log
N(\HI )\geq 13.1$  \citep[``blending in",][]{Parnell88}.   The
Kolmogorov-Smirnov results for the Doppler parameter distribution
for the high and low column density subsamples
would be consistent with blending effects
(\S~\ref{subsec:dopplerparams}).  However, if blending dominates
$dN/dz$, we would expect a general overdensity of absorbers with
redshift, not a localized one. Given the paucity of redshift
density measurements for weak \Lya\ systems  at $z\simlt 0.3$ and
the large cosmic variance we may be seeing  in our sample,  more
high resolution spectra of low $z$ QSOs  are required to determine
the redshift behavior of weak low $z$ \Lya\ forest lines.
Unfortunately,  there are very few
QSOs as bright as
PKS~0405--123 and at 
$0.2<z<1.6$, which permit efficient \Lya\ forest
surveys. 
The Cosmic
Origins Spectrograph, if ever launched,
should do much to address this issue.

\subsection{Clustering and voids}
\label{sec:discussion-clustersvoids}

The most complete work on clustering in the low redshift \Lya\
forest to date is from the $\sim~41$~\kms\ resolution study of
\citet{Penton04}, who find a  two point correlation function
signal  of $\xi(\Delta v<190~{\rm km\,s}^{-1})\sim 3.3$  at the
$4.5\sigma$ significance level  and  $\xi(\Delta v<260\,{\rm
km\,s}^{-1})\sim 2.8$ at  $5.6\sigma$ significance for rest
equivalent width $W_0\geq 65$~m\AA\ ($\log \NHI\simgt 13.1$ for
$b = 25$~\kms ).  They also find a low statistical excess
($3\sigma$) at $260<\Delta v<680$~\kms , as well as a deficit at
$700<\Delta v<4000$~\kms .  Our clustering signal  of $\xi(\Delta
v<250~{\rm km\,s}^{-1}) = 1.2$ is weaker in comparison.  However,
\citet{Janknecht02} found no \Lya -\Lya\ correlations in STIS+VLT
echelle data of HE~0515-4414 ($0.9<z<1.7$), using 235 \Lya\ lines
with $\log \NHI \geq 13.1$. At higher redshifts, results have
been mixed.  A few examples include weak clustering, $\xi(\Delta
v<250~{\rm km\,s}^{-1}) = 2.4$, found by \citeauthor{Rauch92} for
$b<20$~\kms\ at $z\sim 2.8$, albeit at $2.6\sigma $ significance. 
\citet{Rollinde03} also found correlations in the \Lya\ forest
toward pairs of QSOs  at $z\sim 2$ on scales of $\Delta v \simlt
200$~\kms\ in {\it both} the transverse and line of sight
directions with $R=1400$ spectra, using a pixel opacity method.
\cite{Kim01} found  $\xi(\Delta v<100~{\rm km\,s}^{-1})=0.4\pm
0.1$ for  $\log \NHI \geq 12.7$ at $1.5<z<2.4$, with an
increase in clustering at lower redshift and (marginally) at
higher column density. \citet{Cristiani95} also found  $\xi(\Delta v<350~{\rm
km\,s}^{-1}) = 0.34\pm 0.06$ for $\log \NHI \geq 13.3$ at
$2.96<z<3.54$, with clustering strength increasing with column
density.

\citeauthor{Penton04} attribute the excess clustering they found
to filaments at $\Delta v<1000$~\kms\ and the deficit for larger
$\Delta v$ to the presence of voids, analogous to similar behavior
in the galaxy two point correlation function.  There is marginal
evidence for  such larger scale structure in our correlation
function in the form of a $\sim 2.9\sigma$ overdensity at 
$2000<\Delta v<2250$~\kms , which may reflect beating among
substructures in the \Lya\ forest distribution.  A larger sample size is
necessary to determine the reality of such features.

More work has been done on the topic of low redshift \Lya\
absorbers in known galaxy voids \citep[e.g.][and references
therein]{McLin02} than on voids in the low redshift \Lya\ forest
itself. 
The size of the void in the \Lya\ forest  we detect at
$0.0320<z<0.0814$ ($\Delta v = 14000$ \kms , 206$h^{-1}$ comoving 
Mpc) for strong absorbers is on the order of twice the mean size
of large voids as traced by $R\geq 1$ Abell/ACO galaxy clusters
\citep{Stavrev00}, and of the same order in the case of voids in
our weak absorber survey.  Voids in the \Lya\ forest are rare, and
the most widely-known comparable examples with similar \HI\ column
density limits to the ones in our sample are at high redshift.
\citet{Srianand96} found a void toward Tol~1037--2704 at
$2.16286<z<2.20748$  ($59.3h^{-1}$~comoving Mpc) for a rest
equivalent width limit of 0.1~\AA\ ($\log \NHI\approx 13.4$).
The ``Dobrzycki-Bechtold Void" \citep{Dobrzycki91}, for which
\citet{Heap00}  profile-fitted Keck HIRES spectra, has a void at
$3.1513<z<3.2044$ toward Q~0302--003 for logNHI$\leq 13.42$
($47.9h^{-1}$~comoving Mpc).  ``Crotts' Gap" \citep{Crotts87}
which is at $2.5552<z<2.5931$ toward Q~0420--388 for $\log \NHI
\simlt 13.5$, has a depth of $42h^{-1}$~comoving Mpc; 
\citet{Rauch92} found weak lines in Crotts' Gap, but confirm a
significant absorber deficit in it.  \citet{Kim01}  found evidence
for three voids at $1.57 < z < 2.22$ for $\log \NHI \geq
13.5$ toward HE~0505--4414 and HE~2217--2818 of
$50-68h^{-1}$~comoving Mpc, with chance probabilities
$0.01<P<0.05$.  Given the expected column density  {\it vs.}
density perturbation evolution between $z\sim 2$ and $z\sim 0.2$,
the column density levels for which we find our low redshift
void(s) toward PKS~0405--123 ($13.1 \leq \log \NHI \leq 13.3$)
are comparable to density perturbations $\log \rho / \bar{\rho}
\approx 0.6-0.7$, which correspond to $\log \NHI \approx 14.0$
at $z=2$ \citep{Dave99}.  Our \Lya\ forest sample therefore
probes larger density perturbations than these high redshift
studies, and so it is 
reasonable that the voids
we find at low redshift with low column density systems are larger
than the voids found at $z\simgt 1.5-2$, without even factoring in
the growth of voids themselves.

Voids are expected to grow at least as fast as the Hubble flow, 
so the \Lya\ forest void structure toward PKS~0405--123  is at
least consistent with the sizes of higher redshift voids.   At
high redshift, voids are sometimes attributed to ionization from
QSOs or AGN in the plane of the sky (or the ``transverse proximity
effect") \citep[e.g.][]{Liske01,Jakobsen03}. A similar effect may
occur for these low redshift voids. Although we only find one
galaxy in the \Lya\ forest voids from our $\sim 8$~arcmin CFHT 
spectroscopic field, within one degree (1.4-7.4 local frame Mpc)
there are a two galaxies at $z\approx 0.033$ and an AGN at
$z=0.0789$, and a galaxy and AGN at $z\approx 0.12$
(Table~\ref{tab:AGN}).  Because the UV background is expected
to be low at low redshift \citep{Dave99}, 
it is plausible that AGN could
have measurable effects on the \Lya\ forest density over
such distances.

\subsection{Ly$\alpha$ absorber -- galaxy correlations}

We find that $\xi_{Ly\alpha G}$
for $13.5\simlt \log \NHI\simlt 14.0$ is nearly as strong as
the galaxy-galaxy correlation $\xi_{GG}$ 
for the mean luminosity of our
galaxy sample, at least on scales of $200 \simlt \Delta v \simlt
500$~\kms .  
The relative strength of the galaxy-absorber to galaxy-galaxy correlations 
contrasts with previous studies of low redshift \Lya\ absorber --
galaxy correlations \citep[e.g.]{Impey99,Penton02} which find $\xi_{Ly\alpha G}<\xi_{GG}$.
However,
a recent comparison of HIPASS galaxies with $\log \NHI < 15$ absorbers
shows
$\xi_{Ly\alpha G}\approx \xi_{GG}$, integrating over $\sigma$ (the projected
separation) in her
nomenclature over a distance of 1~Mpc 
\citep[][and Ryan-Weber 2005, private communication]{Ryanweber05}. 
ELGs  dominate 
the HIPASS sample \citep[$>80$\%,][]{Doyle05,Meurer05}, as they do ours.
A related
study of the PKS~0405--123 field \citep{Chen05}, with a larger,
more homogeneous galaxy sample, also exhibits a comparable correlation strength
between emission line galaxies (ELGs) and absorbers.  
\citeauthor{Chen05} additionally found a similar relationship between minimum
\HI\ column density and correlation strength, as in this work.
Qualitatively, $\xi_{Ly\alpha G}<\xi_{GG}$ is
expected  from simulations where Ly-a forest absorbers arise in diffuse filaments
that are at lower overdensities and hence less clustered than galaxies.   
Our
result of $\xi_{Ly\alpha G}\sim \xi_{GG}$ would be  consistent with  absorbers of
$13.5\simlt \log \NHI \simlt 14.0$ arising in association with masses similar to
the galaxies in our sample.  
Unfortunately, due to the 
inhomogeneity of our galaxy sample, it is difficult  to make firm quantitative
comparisons to models or other  observations.

The galaxy luminosities in our sample and the similarity in galaxy-absorber
{\it vs.} galaxy-galaxy correlations allow us to constrain the halo masses
associated with \Lya\ absorbers with $\log \NHI \simgt 13.5$.
The mean absolute $r$ magnitude
$\langle M_r\rangle =-19.8 \pm 1.7$ corresponds to 
$\log M/M_\odot \sim 11.3^{+1.1}_{-0.7}$ 
\citep{Berlind03}. 
If we exclude the faintest dwarf galaxy in our sample (no. 31, $z=0.0234$) 
in the magnitude
statistics, which is 3.3 orders of magnitude fainter than the next faintest
galaxy (no. 72) and $>900$~\kms\ from any \Lya\ absorber,
then
$\langle M_r\rangle =-19.9 \pm 1.3$ and the mass constraint is narrowed to 
$\log M/M_\odot \sim 11.3^{+1.0}_{-0.6}$.  This mass range is
intriguingly consistent with those of \Mgii\ absorbers with rest equivalent width
$W_r\geq 1.0$~\AA\ and $0.4<z<0.8$, which have a cross-correlation
length with galaxies of $\sim 5$~Mpc \citep{Bouche04}.
However, there are only
two \Mgii\ absorbers found toward PKS~0405--123 to $W_r=0.2$ and $z<0.41$ 
\citep{Spinrad93},
whereas there are 28 \Lya\ absorbers in the strong sample with $\log \NHI \geq 13.5$.
We thus infer
that the fraction of halo masses in that range associated with \MgII\ compared to \HI\
absorption is on the order of 10\% , barring strong evolutionary effects 
between $0<z<0.4$ and $0.4<z<0.8$.
Photometry in the $rg$ bands is only available for 9 galaxies
in our sample at $z<0.43$, from \citet{Spinrad93} and \citet{Ellingson94};
$\langle g-r \rangle = 0.8\pm 0.3$ which would make typical ages in that galaxy
subsample to be 6.75~Gyr (\citeauthor{Berlind03}).


Accepting that our spectroscopically identified 
galaxy counts are incomplete with a
a non-uniform spatial and redshift selection function,
we confirm over $0<z<0.4$ 
at $\approx 99.0$\% confidence 
the correlation between \HI\ column
density and local galaxy density  found by \citet{Bowen02} at
$z\simlt 0.01$, with a difference that four galaxies fainter than $M_r=-17.5$ are
included in our sample.  The strongest correlation between \HI\
column density and local galaxy counts arises for binning $\Delta
z = 0.02$, which corresponds to $4000<\Delta v < 6000$~\kms ,
larger than the $\pm 1000$~\kms\ which \citeauthor{Bowen02}
found.  This may be an evolutionary effect. More likely, the
larger scale we find could simply reflect the coarse galaxy sample
with which we must work. Better galaxy statistics address
this question \citep{Chen05}. We note that the
$N(\HI )$--galaxy count correlation we find is {\it not} dominated
by metal or \OVI\ systems, because there are only six in our
redshift sensitivity range, and they occur in only 4/17 redshift
bins (one of which happens to contain no galaxies at all, in the
case of the $\Delta z = 0.02$ binning). In particular, the bins
with small galaxy counts and low $N(\HI )$ sums also correlate,
and the effect is {\it not} due to the limited volume sampled for
galaxies at low redshift.

If the trend continues, we would predict a relatively low \HI\
column density at $z\approx 0.45$ and a relatively high $N(\HI )$
at $z\approx 0.43,0.51$, in the vicinity of the highest redshift
\OVI\ absorber.   We already know that there are high $N(\HI )$
absorbers at $z=0.495$ and 0.538. A more complete galaxy census
would permit a more quantitative study between $\log \NHI$ and
local galaxy density, and provide an excellent uniform data set to
study the relationship between galaxy luminosity, impact parameter
and \HI\ column density at low redshift.  Such data would enable
us to disentangle the relationship between galaxies and absorbers
over most of the age of the universe, from $z>3$ 
\citep[e.g.][]{Adelberger03} to the present epoch.  The redshift
and brightness of PKS~0405--123, and the rich structure its sight
line probes, make it an important object for such studies.

\section{Conclusions}
\label{sec:conclusions}

We have performed an analysis of the \Lya\ forest toward
PKS~0405--123 and its relation to a galaxy sample within a 5~arcmin
field.

1. We present STIS E140M echelle data for PKS~0405--123, and
performed profile fits  or measured apparent optical depths for
all of the detected \Lya\ forest over $0<z<0.423$,  plus 
intervening metal and Galactic metal absorption systems.  We
analyzed simulated spectra to determine our sensitivity to line
detections and resolution of close pairs. We created two samples,
a strong one for  $\log \NHI \geq 13.3$ covering
$0.002<z<0.423$ and containing 60 absorbers, and a weak one for
$\log \NHI \geq 13.1$ covering $0.020<z<0.234$ with 44
absorbers. 
Seven absorbers contain metals, all of which show \OVI\ which tend to show
velocity offsets from the corresponding \HI\ absorbers.

2. The Doppler parameter distribution for the strong sample has 
mean and standard deviation $\langle b\rangle =47 \pm 22$~\kms .
For the weak sample, the values are $\langle b\rangle =44 \pm
21$~\kms .  Comparison with analysis from simulated spectra
indicates that the means are inflated,
at least partly due to
line blending and our limited signal-to-noise ratio.

3. The redshift density for the absorbers with column densities
$\log \NHI> 14.0$  is consistent with previous, lower resolution studies. 
For absorbers with $13.1< \log \NHI< 14.0$, the
redshift density is higher than comparable low redshift
studies at the same resolution.   We split the weak sample into low
and high redshift halves, and conclude that our results are
consistent with the values from the literature for $z<0.127$.
However, it appears that cosmic variance produces a high redshift
density at $0.127<z<0.234$.

4. We find evidence of clustering in the \Lya\ forest for $\log
N(\HI ) \geq 13.3$ with a probability of occurrence from
unclustered data of $P<0.005$, with a correlation strength of
$\xi(\Delta v)=1.2$ for velocity differences 
$\Delta v < 250$~\kms .  
The clustering strength 
and scale are consistent with numerical
models for the growth of density perturbations producing the \Lya\
forest.

5. We find evidence of a void at $0.0320<z<0.0814$ in the strong
absorber sample with a probability of random occurrence from
unclustered data 
$P=0.0004$. In the weak sample, there is a void at
$0.0320<z<0.0590$ ($P=0.003$).

6. We cross-correlated the \Lya\ forest samples with an
inhomogeneous survey of  galaxies in the field, using data from a
multi-slit survey from the CFHT and from the literature. We find a
correlation which is maximally significant for absorbers with
$\log \NHI\simgt 13.5-14.0$ over a transverse distance of 1.6~$h^{-1}$~Mpc in 
the local frame, and which is nearly of the same strength
as the galaxy-galaxy correlation for a luminosity equal to the
mean in our galaxy sample. The correlation strength rises for higher
minimum \HI\ column density thresholds; it becomes insignificant
if the maximum column density is limited for a sample of $13.1 \leq \log \NHI \geq 14.3$.
Lower column density
absorbers (with minimum threshold  $13.1 \log \NHI< 13.2$) appear
to have a smaller correlation length with galaxies,
corresponding to velocity differences of
$\Delta v < 125$~\kms .  Higher column density
systems show correlations with galaxies out to $\Delta v <
250$~\kms , which is consistent with higher column density absorbers
arising from larger density
perturbations.  The mean luminosity of the galaxies in our sample
correspond to $\log M/M_\odot = 11.3^{+1.0}_{-0.6}$, which we
take to be representative of the \Lya\ system masses for $13.5 \simlt 
\log \NHI \simlt 14.0$,
given the
similarity in correlation strength.

7. There is a correlation between local summed \HI\ column density
and galaxy counts over the entire strong sample, with a peak
significance on scales of $4000-6000$~\kms\ and a probability of
occurrence from uncorrelated data of $P\approx 
0.010$. Based on this correlation, we predict
a low \HI\ column density at $z\approx 0.45$ and high values at 
$z\approx 0.43$ and $z\approx 0.51$.  

9. A more complete galaxy survey would be necessary to make a more
quantitative study between absorbers and galaxy environment.
PKS~0405--123 provides good illumination along an extended path
through the low redshift \Lya\ forest and galaxy environment.

\acknowledgements

This work was partially supported by the STIS IDT through the
National Optical Astronomical Observatories and by the Goddard
Space Flight Center. Based on observations obtained with the
NASA/ESA Hubble Space Telescope, which is operated by the
Association of Universities for Research in Astronomy, Inc., under
NASA contract NAS 5-26555.  We also used the NASA Extragalactic
Database (NED).  We thank C. Howk for informative discussions
and help with the PKS~0405--123
{\it FUSE} data,
D. Bowen, J. Loveday, E. Ryan-Weber and 
E. Williger for useful
discussions, H. Yee for contributions 
to observing at the CFHT and 
producing the photometric and spectroscopic galaxy catalogue,
and J. Prochaska, T-S. Kim and E. Janknecht for
discussions and assistance in assembling comparison data from the
literature. We also thank the anonymous referee for helpful comments.
TMT received additional support from NASA through LTSA grant NNG
04GG73G.


\clearpage
\begin{deluxetable}{lccrrrcl}
\tablewidth{0pc}
\tablecaption{Absorption lines toward PKS~0405-123
\label{tab:linelist}}
\tablehead{ 
\colhead{species} & \colhead{$z$} & \colhead{error\tablenotemark{a}}     &\colhead{$b$ (\kms )} & \colhead{error\tablenotemark{a}} 
& \colhead{$\log N$} & \colhead{error\tablenotemark{a}} 
& \colhead{lines (\AA )/remarks\tablenotemark{b,c}}} 
\footnotesize
\startdata
	   C II  & 	      -0.000242 & \nodata  &\nodata &\nodata &$>$14.19 &\nodata  &    	   1334\tablenotemark{d}   \\
	   O I   & 	      -0.000137 & \nodata  &\nodata &\nodata &$>$14.43 &\nodata  &    	   1302\tablenotemark{d}   \\
	   AlII  & 	      -0.000100 &\nodata   &\nodata &\nodata &$>$12.82  &\nodata  &    	   1670\tablenotemark{d}   \\
	   FeII  & 	      -0.000098 & 0.000008 &   10.4 &    3.4 & 13.69  &   0.10  &    	   1144,1260,1608   \\
	   N I   & 	      -0.000097 &\nodata   &\nodata &\nodata &$>$13.59  &\nodata  &    	   1200\tablenotemark{d}   \\
	   SiIV  & 	      -0.000081 & 0.000012 &   13.9 &    4.6 & 12.78  &   0.13  &    	   1393,1402   \\
	   SiII  & 	      -0.000078 & 0.000005 &   16.7 &    1.5 & 13.69  &   0.04  &    	   1190,1193,1260,1304,1526   \\
	   SiIII & 	      -0.000013 &\nodata   &\nodata &\nodata &$>$13.65  &\nodata  &    	   1206\tablenotemark{d}   \\
	   C IV  & 	       0.000005 & 0.000005 &   35.4 &    2.2 & 14.14  &   0.03  &    	   1548,1550   \\
	   AlII  & 	       0.000027 &\nodata   &\nodata &\nodata &$>$12.77  &\nodata  &    	   1670\tablenotemark{d}   \\
	   S II  & 	       0.000032 & 0.000005 &    7.2 &    1.2 & 15.15  &   0.14  &    	   1250,1253,1259   \\
	   C II  & 	       0.000042 & \nodata  &\nodata &\nodata &$>$14.80 &\nodata  &    	   1334\tablenotemark{d}   \\
	   SiIV  & 	       0.000056 & 0.000006 &   19.4 &    2.6 & 13.37  &   0.04  &    	   1393,1402   \\
	   P II  & 	       0.000060 &\nodata   &\nodata &\nodata &$>$13.80 &\nodata  &    	   1152\tablenotemark{d,e}   \\
	   FeII  & 	       0.000065 & \nodata  &\nodata &\nodata &$>$14.59  &\nodata  &    	   1608\tablenotemark{d}   \\
	   SiII  & 	       0.000068 &  \nodata &\nodata &\nodata&$>$14.50  &\nodata  &    	   1526\tablenotemark{d}   \\
	   C I   & 	       0.000073 & 0.000006 &   23.4 &    2.4 & 13.87  &   0.04  &          $\lambda \lambda$1157,1277,1280, \\
	         &                      &          &        &        &        &         &              1328,1560,1656                    \\
	   N I   & 	       0.000075 &\nodata   &\nodata &\nodata &$>$14.90  &\nodata  &    	   1200\tablenotemark{d}   \\
	   C II* & 	       0.000077 & \nodata  &\nodata &\nodata &$>$14.56 &\nodata  &    	   1335\tablenotemark{d}   \\
	   NiII  & 	       0.000093 & 0.000004 &    9.4 &    1.8 & 13.39  &   0.06  &    	   1317,1370   \\
	   S II  & 	       0.000097 &  \nodata &\nodata&\nodata&$>$15.73  & \nodata  &    	   1250\tablenotemark{d}   \\
	   O I   & 	       0.000102 & \nodata  &\nodata &\nodata &$>$15.04 &\nodata  &    	   1302\tablenotemark{d}   \\
	   H I   & 	       0.000115 & 0.000032 &\nodata &\nodata & 20.574  &   0.014  &          Ly$\alpha$ \\ 
	   S III & 	       0.000108 & \nodata  &\nodata &\nodata&$>$14.01 & \nodata &          1190 \\
	   H I   & 	       0.002594 & 0.000030 &   28.8 &   16.8 & 13.45  &   0.17  &    	   Ly$\alpha$  \\
	   H I   & 	       0.003838 & 0.000012 &   13.7 &    5.3 & 13.23  &   0.13  &    	   Ly$\alpha$  \\
	   H I   & 	       0.009203 & 0.000014 &   15.6 &    6.9 & 12.83  &   0.14  &    	   Ly$\alpha$  \\
	   H I   & 	       0.011904 & 0.000011 &   18.6 &    5.0 & 13.28  &   0.09  &    	   Ly$\alpha$  \\
	   H I   & 	       0.013920 & 0.000020 &   24.1 &    9.3 & 12.85  &   0.13  &    	   Ly$\alpha$  \\
	   H I   & 	       0.014917 & 0.000010 &   15.0 &    4.3 & 13.02  &   0.09  &    	   Ly$\alpha$  \\
	   H I   & 	       0.015308 & 0.000007 &    9.2 &    3.3 & 12.85  &   0.10  &    	   Ly$\alpha$  \\
	   H I   & 	       0.016810 & 0.000016 &   23.8 &    7.1 & 13.06  &   0.10  &    	   Ly$\alpha$  \\
	   H I   & 	       0.017666 & 0.000008 &   10.2 &    3.4 & 12.75  &   0.10  &    	   Ly$\alpha$  \\
	   H I   & 	       0.018982 & 0.000014 &   16.4 &    6.5 & 12.80  &   0.13  &    	   Ly$\alpha$  \\
	   H I   & 	       0.020342 & 0.000027 &   41.0 &   12.6 & 13.26  &   0.10  &    	   Ly$\alpha$  \\
	   H I   & 	       0.024999 & 0.000013 &   16.7 &    5.6 & 12.93  &   0.11  &    	   Ly$\alpha$  \\
	   H I   & 	       0.029853 & 0.000008 &   10.7 &    3.0 & 13.26  &   0.11  &    	   Ly$\alpha$  \\
	   H I   & 	       0.030004 & 0.000004 &   19.6 &    1.2 & 14.39  &   0.04  &    	   Ly$\alpha , \beta, \delta$  \\
	   H I   & 	       0.031960 & 0.000024 &   55.9 &   11.0 & 13.35  &   0.07  &    	   Ly$\alpha$  \\
	   H I   & 	       0.046643 & 0.000012 &   20.8 &    5.2 & 12.99  &   0.09  &    	   Ly$\alpha$  \\
	   H I   & 	       0.058964 & 0.000035 &   54.5 &   19.6 & 13.26  &   0.12  &    	   Ly$\alpha$  \\
	   H I   & 	       0.059246 & 0.000012 &   15.8 &    5.8 & 12.93  &   0.15  &    	   Ly$\alpha$  \\
	   H I   & 	       0.063663 & 0.000033 &   44.4 &   14.6 & 13.10  &   0.11  &    	   Ly$\alpha$  \\
	   H I   & 	       0.065515 & 0.000058 &   66.1 &   30.2 & 13.11  &   0.15  &    	   Ly$\alpha$  \\
	   H I   & 	       0.067099 & 0.000023 &   24.2 &    9.3 & 12.79  &   0.13  &    	   Ly$\alpha$  \\
	   H I   & 	       0.068155 & 0.000012 &   11.0 &    5.2 & 12.71  &   0.14  &    	   Ly$\alpha$  \\
	   H I   & 	       0.069011 & 0.000032 &   46.0 &   14.8 & 13.05  &   0.11  &    	   Ly$\alpha$  \\
	   H I   & 	       0.071935 & 0.000010 &   12.8 &    4.2 & 12.56  &   0.10  &    	   Ly$\alpha$  \\
	   H I   & 	       0.072182 & 0.000018 &   44.6 &    8.7 & 13.27  &   0.06  &    	   Ly$\alpha$  \\
	   H I   & 	       0.072507 & 0.000008 &   22.5 &    3.6 & 13.20  &   0.06  &    	   Ly$\alpha$  \\
	   H I   & 	       0.073902 & 0.000041 &   48.4 &   25.2 & 13.08  &   0.17  &    	   Ly$\alpha$  \\
	   H I   & 	       0.074487 & 0.000013 &   21.0 &    5.4 & 13.02  &   0.09  &    	   Ly$\alpha$  \\
	   H I   & 	       0.074750 & 0.000020 &   27.5 &    8.5 & 12.98  &   0.11  &    	   Ly$\alpha$  \\
	   H I   & 	       0.075233 & 0.000031 &   55.5 &   13.3 & 13.20  &   0.08  &    	   Ly$\alpha$  \\
	   H I   & 	       0.081387 & 0.000008 &   53.6 &    3.0 & 13.81  &   0.02  &    	   Ly$\alpha$  \\
	   H I   & 	       0.082277 & 0.000022 &   27.3 &    9.8 & 12.88  &   0.12  &    	   Ly$\alpha$  \\
	   H I   & 	       0.084851 & 0.000042 &   45.1 &   20.2 & 13.06  &   0.15  &    	   Ly$\alpha$  \\
	   H I   & 	       0.085655 & 0.000022 &   19.4 &    9.2 & 12.65  &   0.16  &    	   Ly$\alpha$  \\
	   H I   & 	       0.091843 & 0.000004 &   40.0 &    0.9 & 14.54  &   0.02  &    	   Ly$\alpha , \beta , \gamma$  \\
	   H I   & 	       0.093387 & 0.000010 &   17.8 &    3.9 & 13.05  &   0.07  &    	   Ly$\alpha$  \\
	   H I   & 	       0.096584 & 0.000004 &   37.2 &    0.8 & 14.67  &   0.02  &    	   Ly$\alpha ,\beta ,\gamma$  \\
	   H I   & 	       0.097159 & 0.000015 &   24.6 &    5.6 & 13.19  &   0.10  &    	   Ly$\alpha$  \\
	   H I   & 	       0.101450 & 0.000027 &   33.6 &   13.0 & 12.96  &   0.12  &    	   Ly$\alpha$  \\
	   H I   & 	       0.102982 & 0.000032 &   60.8 &   14.1 & 13.35  &   0.08  &    	   Ly$\alpha$  \\
	   H I   & 	       0.107973 & 0.000014 &   11.3 &    5.8 & 12.69  &   0.16  &    	   Ly$\alpha$  \\
	   H I   & 	       0.117241 & 0.000025 &   40.7 &    9.3 & 13.15  &   0.08  &    	   Ly$\alpha$  \\
	   H I   & 	       0.118405 & 0.000042 &   45.4 &   16.0 & 13.00  &   0.13  &    	   Ly$\alpha$  \\
	   H I   & 	       0.119299 & 0.000028 &   35.9 &   12.6 & 13.04  &   0.12  &    	   Ly$\alpha$  \\
	   H I   & 	       0.119579 & 0.000027 &   20.2 &   12.3 & 12.71  &   0.20  &    	   Ly$\alpha$  \\
	   H I   & 	       0.130972 & 0.000014 &   48.6 &    5.2 & 13.46  &   0.04  &    	   Ly$\alpha$  \\
	   H I   & 	       0.132293 & 0.000004 &   20.5 &    1.6 & 13.60  &   0.03  &    	   Ly$\alpha$  \\
	   H I   & 	       0.133064 & 0.000017 &   36.2 &    6.1 & 13.35  &   0.06  &    	   Ly$\alpha$  \\
	   H I   & 	       0.133775 & 0.000021 &   47.0 &    7.8 & 13.37  &   0.06  &    	   Ly$\alpha$  \\
	   H I   & 	       0.136460 & 0.000021 &   49.3 &    8.3 & 13.37  &   0.06  &    	   Ly$\alpha$  \\
	   H I   & 	       0.139197 & 0.000025 &   34.8 &   11.5 & 13.02  &   0.11  &    	   Ly$\alpha$  \\
	   H I   & 	       0.151363 & 0.000020 &   41.5 &    7.3 & 13.26  &   0.06  &    	   Ly$\alpha$  \\
	   H I   & 	       0.152201 & 0.000006 &   22.5 &    2.1 & 13.52  &   0.04  &    	   Ly$\alpha$  \\
	   H I   & 	       0.153044 & 0.000009 &   50.8 &    3.1 & 13.82  &   0.03  &    	   Ly$\alpha$  \\
	   H I   & 	       0.161206 & 0.000017 &   56.6 &    6.2 & 13.73  &   0.04  &    	   Ly$\alpha$  \\
	   H I   & 	       0.161495 & 0.000008 &   18.2 &    3.4 & 13.27  &   0.09  &    	   Ly$\alpha$  \\
	   H I   & 	       0.163016 & 0.000013 &   38.2 &    4.4 & 13.47  &   0.04  &    	   Ly$\alpha$  \\
	   FeIII & 	       0.166216 & 0.000007 &   22.5 &    2.6 & 14.24  &   0.04  &    	   1122   \\
	   H I   & 	       0.166628 & 0.000006 &    9.0 &    2.5 & 13.30  &   0.11  &    	   Ly$\alpha ,\beta$  \\
	   H I   & 	       0.166962 & 0.000032 &  112.2 &    8.3 & 14.17  &   0.06  &    	   Ly$\alpha ,\beta$  \\
	   SiII  & 	       0.166989 & 0.000006 &   11.2 &    2.2 & 12.64  &   0.06  &    	   1190,1193,1260,1304   \\
	   C II  & 	       0.166998 & \nodata  &\nodata&\nodata&$>$14.35  &\nodata  &    	   \tablenotemark{f}   \\
	   O VI  & 	       0.167007 & 0.000019 &   83.2 &    6.7 & 14.81  &   0.03  &    	   1031,1037   \\
	   N III & 	       0.167058 &  \nodata &\nodata&\nodata&$<$14.59  &\nodata  &    	   \tablenotemark{f}   \\
	   SiIII & 	       0.167083 & 0.000020 &   32.8 &    3.6 & 13.08  &   0.09  &    	   1206   \\
	   N V   & 	       0.167116 & 0.000018 &   61.6 &    6.8 & 14.06  &   0.04  &    	   1238,1242   \\
	   SiIV  & 	       0.167132 & 0.000009 &   34.5 &    3.0 & 13.44  &   0.03  &    	   1393,1402   \\
	   O I   & 	       0.167132 & 0.000014 &   18.2 &    5.2 & 13.92  &   0.10  &    	   1302   \\
	   H I   & 	       0.167139 & 0.000004 &   29.4 &    1.0 & 16.45  &   0.05  &    	   \tablenotemark{f}; $b$ from Ly$\alpha ,\beta$  \\
	   C II  & 	       0.167144 & 0.000004 &    8.7 &    1.4 & 13.82  &   0.06  &    	   1334   \\
	   SiII  & 	       0.167147 & 0.000002 &   10.3 &    0.7 & 13.32  &   0.04  &    	   1190,1193,1260,1304   \\
	   N II  & 	       0.167152 & \nodata  &\nodata&\nodata&$>$14.24  &\nodata  &    	   \tablenotemark{f}   \\
	   FeII  & 	       0.167167 & 0.000017 &   21.0 &    6.4 & 13.56  &   0.10  &    	   1144   \\
	   SiIII & 	       0.167177 &\nodata   &\nodata&\nodata&$>$13.39  &\nodata  &    	   \tablenotemark{f}   \\
	   H I   & 	       0.170061 & 0.000027 &   38.2 &   12.4 & 13.07  &   0.10  &    	   Ly$\alpha$  \\
	   H I   & 	       0.170316 & 0.000017 &   17.4 &    6.8 & 12.73  &   0.12  &    	   Ly$\alpha$  \\
	   H I   & 	       0.170522 & 0.000015 &   18.9 &    6.2 & 12.82  &   0.10  &    	   Ly$\alpha$  \\
	   H I   & 	       0.171466 & 0.000019 &   29.1 &    6.9 & 13.10  &   0.09  &    	   Ly$\alpha$  \\
	   H I   & 	       0.171962 & 0.000012 &    7.8 &    4.7 & 12.53  &   0.16  &    	   Ly$\alpha$  \\
	   H I   & 	       0.172244 & 0.000016 &   16.5 &    5.8 & 12.81  &   0.12  &    	   Ly$\alpha$  \\
	   H I   & 	       0.173950 & 0.000048 &   36.2 &   19.4 & 12.80  &   0.18  &    	   Ly$\alpha$  \\
	   H I   & 	       0.178761 & 0.000017 &   57.7 &    5.9 & 13.68  &   0.04  &    	   Ly$\alpha$  \\
	   H I   & 	       0.181032 & 0.000022 &   23.0 &    8.9 & 12.86  &   0.13  &    	   Ly$\alpha$  \\
	   H I   & 	       0.182715 & 0.000004 &   47.5 &    2.0 & 15.07  &   0.09  &    	   Ly$\alpha ,\beta$  \\
	   O VI  & 	       0.182918 & 0.000011 &   26.9 &    8.4 & 14.04  &   0.18  &    	   1031,1037   \\
	   H I   & 	       0.184248 & 0.000019 &   24.6 &    6.7 & 12.89  &   0.10  &    	   Ly$\alpha$  \\
	   H I   & 	       0.184742 & 0.000015 &   18.3 &    5.7 & 12.97  &   0.10  &    	   Ly$\alpha$  \\
	   H I   & 	       0.185082 & 0.000022 &   20.1 &    8.3 & 12.85  &   0.14  &    	   Ly$\alpha$  \\
	   H I   & 	       0.190865 & 0.000032 &   53.0 &   11.4 & 13.25  &   0.08  &    	   Ly$\alpha$  \\
	   H I   & 	       0.192010 & 0.000024 &   18.0 &    9.4 & 12.71  &   0.19  &    	   Ly$\alpha$  \\
	   H I   & 	       0.192550 & 0.000066 &   87.1 &   26.5 & 13.28  &   0.11  &    	   Ly$\alpha$  \\
	   H I   & 	       0.194617 & 0.000042 &   80.6 &   15.4 & 13.37  &   0.07  &    	   Ly$\alpha$  \\
	   H I   & 	       0.195607 & 0.000039 &   36.0 &   14.8 & 12.94  &   0.15  &    	   Ly$\alpha$  \\
	   H I   & 	       0.196783 & 0.000057 &   76.7 &   23.9 & 13.33  &   0.11  &    	   Ly$\alpha$  \\
	   H I   & 	       0.204022 & 0.000021 &   21.8 &    7.8 & 12.89  &   0.12  &    	   Ly$\alpha$  \\
	   H I   & 	       0.204680 & 0.000086 &   42.9 &   39.9 & 12.78  &   0.30  &    	   Ly$\alpha$  \\
	   H I   & 	       0.205024 & 0.000049 &   47.8 &   19.7 & 13.03  &   0.14  &    	   Ly$\alpha$  \\
	   H I   & 	       0.205407 & 0.000031 &   24.5 &   13.1 & 12.84  &   0.18  &    	   Ly$\alpha$  \\
	   H I   & 	       0.207013 & 0.000040 &   46.0 &   16.6 & 13.06  &   0.12  &    	   Ly$\alpha$  \\
	   H I   & 	       0.207443 & 0.000038 &   37.7 &   16.9 & 12.99  &   0.15  &    	   Ly$\alpha$  \\
	   H I   & 	       0.210714 & 0.000015 &   41.7 &    5.3 & 13.43  &   0.05  &    	   Ly$\alpha$  \\
	   H I   & 	       0.211165 & 0.000020 &   29.9 &    6.9 & 13.24  &   0.09  &    	   Ly$\alpha$  \\
	   H I   & 	       0.212973 & 0.000031 &   40.4 &   12.0 & 13.13  &   0.10  &    	   Ly$\alpha$  \\
	   H I   & 	       0.213295 & 0.000012 &    8.9 &    4.6 & 12.69  &   0.15  &    	   Ly$\alpha$  \\
	   H I   & 	       0.214370 & 0.000031 &   25.7 &   11.8 & 12.86  &   0.16  &    	   Ly$\alpha$  \\
	   H I   & 	       0.215037 & 0.000017 &   27.1 &    6.3 & 13.01  &   0.08  &    	   Ly$\alpha$  \\
	   H I   & 	       0.215951 & 0.000007 &   25.0 &    2.4 & 13.49  &   0.04  &    	   Ly$\alpha$  \\
	   H I   & 	       0.216478 & 0.000046 &   45.5 &   22.5 & 12.96  &   0.16  &    	   Ly$\alpha$  \\
	   H I   & 	       0.217052 & 0.000045 &   31.4 &   20.1 & 12.83  &   0.20  &    	   Ly$\alpha$  \\
	   H I   & 	       0.217450 & 0.000028 &   26.6 &   10.6 & 12.84  &   0.14  &    	   Ly$\alpha$  \\
	   H I   & 	       0.217959 & 0.000013 &   16.1 &    4.9 & 12.93  &   0.10  &    	   Ly$\alpha$  \\
	   H I   & 	       0.218465 & 0.000023 &   28.5 &    8.3 & 13.03  &   0.10  &    	   Ly$\alpha$  \\
	   H I   & 	       0.219863 & 0.000036 &   36.0 &   13.1 & 12.99  &   0.13  &    	   Ly$\alpha$  \\
	   H I   & 	       0.221067 & 0.000036 &   44.2 &   13.3 & 13.05  &   0.11  &    	   Ly$\alpha$  \\
	   H I   & 	       0.223574 & 0.000030 &   27.6 &   12.6 & 12.99  &   0.15  &    	   Ly$\alpha$  \\
	   H I   & 	       0.224524 & 0.000023 &   27.3 &    8.5 & 13.05  &   0.11  &    	   Ly$\alpha$  \\
	   H I   & 	       0.225186 & 0.000047 &   41.3 &   19.4 & 12.97  &   0.16  &    	   Ly$\alpha$  \\
	   H I   & 	       0.225671 & 0.000028 &   25.9 &   10.9 & 12.82  &   0.14  &    	   Ly$\alpha$  \\
	   H I   & 	       0.228234 & 0.000011 &   20.1 &    3.7 & 13.24  &   0.06  &    	   Ly$\alpha$  \\
	   H I   & 	       0.228622 & 0.000031 &   18.2 &   11.4 & 12.68  &   0.21  &    	   Ly$\alpha$  \\
	   H I   & 	       0.229402 & 0.000039 &   27.6 &   14.1 & 12.84  &   0.18  &    	   Ly$\alpha$  \\
	   H I   & 	       0.229878 & 0.000039 &   51.5 &   15.3 & 13.23  &   0.10  &    	   Ly$\alpha$  \\
	   H I   & 	       0.233762 & 0.000056 &   58.3 &   21.9 & 13.14  &   0.13  &    	   Ly$\alpha$  \\
	   H I   & 	       0.235193 & 0.000004 &    4.6 &    1.6 & 12.74  &   0.09  &    	   Ly$\alpha$  \\
	   H I   & 	       0.239255 & 0.000067 &   73.5 &   30.7 & 13.27  &   0.14  &    	   Ly$\alpha$  \\
	   H I   & 	       0.239738 & 0.000028 &   19.9 &   10.8 & 12.78  &   0.21  &    	   Ly$\alpha$  \\
	   H I   & 	       0.240542 & 0.000032 &   63.2 &   13.8 & 13.39  &   0.07  &    	   Ly$\alpha$  \\
	   H I   & 	       0.242414 & 0.000020 &   12.3 &    7.6 & 12.67  &   0.19  &    	   Ly$\alpha$  \\
	   H I   & 	       0.242746 & 0.000029 &   18.0 &   10.7 & 12.73  &   0.19  &    	   Ly$\alpha$  \\
	   H I   & 	       0.245130 & 0.000040 &   55.6 &   14.5 & 13.32  &   0.09  &    	   Ly$\alpha$  \\
	   H I   & 	       0.245543 & 0.000007 &   25.4 &    2.3 & 13.72  &   0.04  &    	   Ly$\alpha ,\beta$  \\
	   H I   & 	       0.248579 & 0.000024 &   18.9 &    8.6 & 13.03  &   0.15  &    	   Ly$\alpha$  \\
	   H I   & 	       0.251549 & 0.000017 &   31.9 &    5.6 & 13.32  &   0.07  &    	   Ly$\alpha$  \\
	   H I   & 	       0.252126 & 0.000017 &   15.5 &    6.3 & 12.86  &   0.13  &    	   Ly$\alpha$  \\
	   H I   & 	       0.253303 & 0.000025 &   34.5 &    8.5 & 13.25  &   0.09  &    	   Ly$\alpha$  \\
	   H I   & 	       0.258613 & 0.000022 &   46.6 &    7.4 & 13.45  &   0.06  &    	   Ly$\alpha$  \\
	   H I   & 	       0.260439 & 0.000012 &   33.5 &    3.8 & 13.49  &   0.04  &    	   Ly$\alpha$  \\
	   H I   & 	       0.265150 & 0.000056 &   69.6 &   22.1 & 13.25  &   0.11  &    	   Ly$\alpha$  \\
	   H I   & 	       0.267558 & 0.000014 &   18.3 &    5.0 & 13.06  &   0.09  &    	   Ly$\alpha$  \\
	   H I   & 	       0.267804 & 0.000028 &   22.8 &   10.6 & 12.87  &   0.15  &    	   Ly$\alpha$  \\
	   H I   & 	       0.269300 & 0.000008 &    7.5 &    3.2 & 13.08  &   0.15  &    	   Ly$\alpha$  \\
	   H I   & 	       0.270337 & 0.000040 &   48.2 &   14.6 & 13.10  &   0.10  &    	   Ly$\alpha$  \\
	   H I   & 	       0.270869 & 0.000061 &   72.4 &   26.5 & 13.31  &   0.12  &    	   Ly$\alpha$  \\
	   H I   & 	       0.271375 & 0.000030 &   31.9 &   10.7 & 13.08  &   0.14  &    	   Ly$\alpha$  \\
	   H I   & 	       0.284522 & 0.000011 &   12.4 &    3.9 & 12.92  &   0.10  &    	   Ly$\alpha$  \\
	   H I   & 	       0.287166 & 0.000007 &    6.3 &    2.3 & 13.08  &   0.12  &    	   Ly$\alpha$  \\
	   H I   & 	       0.288329 & 0.000021 &   27.1 &    7.2 & 13.31  &   0.09  &    	   Ly$\alpha$  \\
	   H I   & 	       0.290418 & 0.000027 &   22.1 &    9.1 & 12.95  &   0.14  &    	   Ly$\alpha$  \\
	   H I   & 	       0.291008 & 0.000019 &   16.6 &    6.9 & 12.98  &   0.14  &    	   Ly$\alpha$  \\
	   H I   & 	       0.294234 & 0.000014 &   11.1 &    4.9 & 12.84  &   0.14  &    	   Ly$\alpha$  \\
	   H I   & 	       0.295227 & 0.000030 &   61.9 &    9.6 & 13.47  &   0.06  &    	   Ly$\alpha$  \\
	   H I   & 	       0.297658 & 0.000009 &   35.2 &    2.5 & 14.00  &   0.03  &    	   Ly$\alpha ,\beta ,\gamma $  \\
	   H I   & 	       0.299042 & 0.000028 &   53.4 &    9.7 & 13.38  &   0.07  &    	   Ly$\alpha$  \\
	   H I   & 	       0.299953 & 0.000067 &   72.6 &   26.4 & 13.32  &   0.12  &    	   Ly$\alpha$  \\
	   H I   & 	       0.304958 & 0.000060 &   50.8 &   24.6 & 13.23  &   0.16  &    	   Ly$\alpha$  \\
	   H I   & 	       0.306748 & 0.000046 &   55.2 &   19.3 & 13.30  &   0.11  &    	   Ly$\alpha$  \\
	   H I   & 	       0.307191 & 0.000009 &    7.0 &    3.2 & 12.75  &   0.13  &    	   Ly$\alpha$  \\
	   H I   & 	       0.309849 & 0.000029 &   27.5 &   11.2 & 12.98  &   0.13  &    	   Ly$\alpha$  \\
	   H I   & 	       0.311721 & 0.000042 &   50.2 &   15.7 & 13.28  &   0.10  &    	   Ly$\alpha$  \\
	   H I   & 	       0.312185 & 0.000018 &   18.0 &    5.8 & 13.03  &   0.11  &    	   Ly$\alpha$  \\
	   H I   & 	       0.312952 & 0.000081 &   45.1 &   25.1 & 13.14  &   0.24  &    	   Ly$\alpha$  \\
	   H I   & 	       0.313341 & 0.000073 &   43.1 &   21.9 & 13.17  &   0.22  &    	   Ly$\alpha$  \\
	   H I   & 	       0.313822 & 0.000088 &   71.1 &   34.1 & 13.35  &   0.16  &    	   Ly$\alpha$  \\
	   H I   & 	       0.320076 & 0.000011 &   26.0 &    3.4 & 13.53  &   0.05  &    	   Ly$\alpha$  \\
	   H I   & 	       0.321160 & 0.000043 &   51.7 &   15.3 & 13.30  &   0.10  &    	   Ly$\alpha$  \\
	   H I   & 	       0.325046 & 0.000035 &   81.4 &   11.0 & 13.67  &   0.05  &    	   Ly$\alpha$  \\
	   H I   & 	       0.327277 & 0.000033 &   52.3 &   11.1 & 13.48  &   0.08  &    	   Ly$\alpha$  \\
	   H I   & 	       0.328432 & 0.000024 &   20.5 &    8.4 & 13.04  &   0.14  &    	   Ly$\alpha$  \\
	   H I   & 	       0.328982 & 0.000014 &   13.1 &    4.7 & 12.90  &   0.11  &    	   Ly$\alpha$  \\
	   H I   & 	       0.334017 & 0.000008 &   35.7 &    2.4 & 13.85  &   0.03  &    	   Ly$\alpha ,\beta$  \\
	   H I   & 	       0.335821 & 0.000040 &   47.0 &   14.7 & 13.11  &   0.10  &    	   Ly$\alpha$  \\
	   H I   & 	       0.336713 & 0.000015 &   19.5 &    5.2 & 12.87  &   0.09  &    	   Ly$\alpha$  \\
	   H I   & 	       0.337215 & 0.000011 &   11.4 &    3.7 & 12.79  &   0.10  &    	   Ly$\alpha$  \\
	   H I   & 	       0.339624 & 0.000059 &   41.5 &   26.1 & 13.03  &   0.20  &    	   Ly$\alpha$  \\
	   H I   & 	       0.341860 & 0.000025 &   44.4 &    7.5 & 13.59  &   0.07  &    	   Ly$\alpha$  \\
	   H I   & 	       0.342339 & 0.000036 &   49.8 &   11.8 & 13.50  &   0.09  &    	   Ly$\alpha$  \\
	   H I   & 	       0.346451 & 0.000017 &   18.6 &    5.4 & 13.16  &   0.10  &    	   Ly$\alpha$  \\
	   H I   & 	       0.348430 & 0.000056 &   28.8 &   19.2 & 13.09  &   0.22  &    	   Ly$\alpha$  \\
	   H I   & 	       0.350988 & 0.000008 &   40.0 &    2.4 & 14.20  &   0.03  &    	   Ly$\alpha ,\beta ,\gamma$  \\
	   H I   & 	       0.351511 & 0.000012 &   27.3 &    3.8 & 13.56  &   0.05  &    	   Ly$\alpha ,\beta$  \\
	   H I   & 	       0.352119 & 0.000015 &   35.1 &    4.8 & 13.60  &   0.05  &    	   Ly$\alpha ,\beta$  \\
	   C IV  & 	       0.360757 & 0.000009 &    8.3 &    3.0 & 13.76  &   0.16  &    	   1548,1550\tablenotemark{g}   \\
	   C III & 	       0.360770 &\nodata   &\nodata&\nodata&$>$13.78  & \nodata &    	   \tablenotemark{f}   \\
	   H I   & 	       0.360799 & 0.000002 &   22.6 &    0.6 & 15.21  &   0.02  &    	   Ly$\alpha ,\beta ,\gamma ,\delta ,\epsilon ,\zeta ,\eta ,\theta $  \\
	   SiIII & 	       0.360816 & 0.000014 &   19.3 &    4.2 & 12.72  &   0.08  &    	   1206   \\
	   H I   & 	       0.361025 & 0.000048 &  134.1 &   17.2 & 14.35  &   0.04  &    	   Ly$\alpha ,\beta $  \\
	   O VI  & 	       0.361560 & 0.000013 &   28.7 &   14.4 & 13.93  &   0.36  &    	   1031,1037   \\
	   O VI  & 	       0.363346 & 0.000004 &    6.7 &    1.8 & 13.52  &   0.09  &    	   1031,1037   \\
	   H I   & 	       0.363425 & 0.000016 &   30.0 &    5.1 & 13.59  &   0.06  &    	   Ly$\alpha$  \\
	   H I   & 	       0.372397 & 0.000031 &   21.7 &   10.2 & 12.97  &   0.16  &    	   Ly$\alpha$  \\
	   H I   & 	       0.376324 & 0.000024 &   15.3 &    7.9 & 13.08  &   0.17  &    	   Ly$\alpha$  \\
	   H I   & 	       0.380809 & 0.000051 &   62.2 &   19.1 & 13.38  &   0.10  &    	   Ly$\alpha$  \\
	   H I   & 	       0.385367 & 0.000030 &   30.6 &    9.3 & 13.23  &   0.11  &    	   Ly$\alpha$  \\
	   H I   & 	       0.386463 & 0.000013 &    9.8 &    4.5 & 12.96  &   0.14  &    	   Ly$\alpha$  \\
	   H I   & 	       0.386720 & 0.000017 &   19.3 &    5.3 & 13.25  &   0.10  &    	   Ly$\alpha$  \\
	   H I   & 	       0.395816 & 0.000050 &   51.3 &   15.5 & 13.34  &   0.11  &    	   Ly$\alpha$  \\
	   H I   & 	       0.399031 & 0.000053 &   56.6 &   18.4 & 13.34  &   0.11  &    	   Ly$\alpha$  \\
	   H I   & 	       0.400105 & 0.000049 &   52.9 &   17.3 & 13.30  &   0.11  &    	   Ly$\alpha$  \\
	   H I   & 	       0.405706 & 0.000004 &   32.8 &    0.8 & 14.99  &   0.02  &    	   Ly$\alpha ,\beta ,\gamma ,\delta ,\epsilon ,\zeta ,\eta ,\theta ,\kappa$  \\
	   H I   & 	       0.408875 & 0.000009 &   37.3 &    2.1 & 14.36  &   0.03  &    	   Ly$\alpha ,\beta ,\gamma ,\delta ,\epsilon ,\zeta$  \\
	   H I   & 	       0.409563 & 0.000017 &   24.6 &    5.0 & 13.52  &   0.08  &    	   Ly$\alpha$  \\
	   H I   & 	       0.418175 & 0.000036 &   45.1 &   11.1 & 13.38  &   0.09  &    	   Ly$\alpha ,\beta$  \\
	   C III & 	       0.495087 & 0.000007 &   12.0 &    1.9 & 13.18  &   0.06  &    	    977   \\
	   H I   & 	       0.495112 & 0.000038 &   61.8 &   13.1 & 14.39  &   0.07  &    	   Ly$\beta ,\gamma$  \\
	   O VI  & 	       0.495122 & 0.000013 &   44.9 &    3.1 & 14.49  &   0.03  &    	   1031,1037   \\
	   O IV  & 	       0.495122 &\nodata   &\nodata&\nodata&$>$14.34  & \nodata &    	   \tablenotemark{f}   \\
	   H I   & 	       0.538301 & 0.000014 &   22.5 &    3.7 & 14.22  &   0.06  &    	   Ly$\beta ,\gamma$  \\
\enddata
\vspace{2mm}
\tablenotetext{a}{Errors are 1~$\sigma$, and assume that the component structure is correct.}
\tablenotetext{b}{\phantom{.}$\!$Lines used in profile fits: $\lambda \lambda$ denotes a doublet, $\lambda \lambda \lambda$ a triplet.
Some $\geq 4\sigma$ features in the STIS data, in particular blueward of $\sim 1200$~\AA , 
correspond to higher order Lyman lines which did not provide useful
constraints for profile fits.}
\tablenotetext{c}{We find unidentified features at 1154.93, 
1168.96, 1172.53, 1180.10, 1203.32 and 1404.90~\AA .  The FUSE
data do not confirm the ones at $\lambda < 1187$~\AA , however.}
\tablenotetext{d}{Upper limits from apparent optical depth method of \citet{Savage91}.}
\tablenotetext{e}{FUSE data used for \ion{P}{2} upper limit column density.}
\tablenotetext{f}{Column density from \citet{Prochaska04}.}
\tablenotetext{g}{The \CIV\ fit is from archival STIS G230M data.}
\end{deluxetable}

\clearpage
\begin{deluxetable}{ccccccc}
\tablewidth{0pc}
\tablecaption{Statistics of simulated data: Doppler parameters
\label{tab:simulations_b}}
\tablehead{ 
\colhead{$snr$\tablenotemark{a}} & \multicolumn{2}{c}{Doppler parameter (\kms )} & \multicolumn{2}{c}{Doppler parameter (\kms )} & \multicolumn{2}{c}{Doppler parameter (\kms )} \\
\colhead{}    & \colhead{input\tablenotemark{b}} & \colhead{recovered\tablenotemark{c}} & \colhead{input\tablenotemark{b}} 
& \colhead{recovered\tablenotemark{c}} & \colhead{input\tablenotemark{b}} & \colhead{recovered\tablenotemark{c}}    }
\startdata  
\phn 4        & $17.5\pm 0.3$	& $22.1\pm 7.3(5.4)$  & $24.7\pm 0.2$	&  $29.3\pm 9.1(7.4)$ & $35.0\pm 0.3$	& $36.8\pm 11.3(\phn 9.0)$  \\
\phn 7        & $17.5\pm 0.3$	& $19.9\pm 5.9(5.3)$  & $24.7\pm 0.3$	&  $28.3\pm 8.4(7.5)$ & $35.0\pm 0.3$	& $39.7\pm 11.8(10.5)$ \\
12            & $17.5\pm 0.3$	& $19.0\pm 6.3(5.3)$  & $24.7\pm 0.3$	&  $26.4\pm 8.6(7.4)$ & $35.0\pm 0.3$	& $35.8\pm 12.5(10.2)$ \\
\enddata
\vspace{2mm}
\tablenotetext{a}{Signal to noise ratio per pixel in simulated STIS spectrum.}
\tablenotetext{b}{Mean and first order about the mean for input Doppler parameters in simulations
{\it which were recovered}.}
\tablenotetext{c}{Mean and first order about the mean for recovered Doppler parameters in simulations,
   with the mean in the formal $1\sigma$ profile fitting errors in parentheses.  The mean recovered Doppler
   parameters are all within the mean profile fitting errors of the input values.}
\end{deluxetable}

\clearpage
\begin{deluxetable}{rccc}
\tablewidth{0pc}
\tablecaption{Statistics of simulated data: \HI\ column densities
\label{tab:simulations_nhi}}
\tablehead{ 
\colhead{$snr$\tablenotemark{a}} & \colhead{Doppler\tablenotemark{b}}  & \multicolumn{2}{c}{$\log N(\HI )$}  \\
\colhead{}    & \colhead{parameter (\kms )} & \colhead{input\tablenotemark{c}} & \colhead{recovered\tablenotemark{d}}   }
\startdata  
4         & 17.5 &    $13.11\pm 0.10$ & $13.28\pm 0.10(0.09)$  \\
7         & 17.5 &     $12.86\pm 0.09$ & $12.94\pm 0.12(0.09)$  \\
12        & 17.5 &     $12.59\pm 0.09$ & $12.66\pm 0.10(0.10)$  \\
&&\\
4         & 24.7 &     $13.15\pm 0.10$ & $13.32\pm 0.12(0.09)$  \\
7         & 24.7 &     $12.91\pm 0.10$ & $13.03\pm 0.11(0.09)$  \\
12        & 24.7 &     $12.68\pm 0.09$ & $12.74\pm 0.11(0.10)$  \\
&&\\
4         & 35.0 &     $13.21\pm 0.11$ & $13.37\pm 0.13(0.09)$ \\
7         & 35.0 &     $12.98\pm 0.10$ & $13.09\pm 0.11(0.09)$ \\
12        & 35.0 &     $12.67\pm 0.10$ & $12.76\pm 0.11(0.10)$ \\
\enddata
\vspace{2mm}
\tablenotetext{a}{Signal to noise ratio per pixel in simulated STIS spectrum.}
\tablenotetext{b}{Simulation sets are specified by input Doppler parameter.}
\tablenotetext{c}{Mean and first order about the mean for input log \HI\ column densities in simulations
{\it which were recovered}.}
\tablenotetext{d}{Mean and first order about the mean for recovered log \HI\ column densities in simulations,
   with the mean in the formal $1\sigma$ profile fitting errors in parentheses.  }
\end{deluxetable}

\clearpage
\begin{deluxetable}{cccccccl}
\tablewidth{0pc}
\tablecaption{\Lya\ forest pairs with $\Delta v < 250$~\kms\ and $\log N(\HI ) \geq 13.3$
\label{tab:lyapairs}}
\tablehead{ 
\colhead{$z_1$} & \colhead{$z_2$} & \colhead{$\Delta v$} & \colhead{$\log N(\HI )_1$}     &\colhead{$\log N(\HI )_2$}& \colhead{$b_1$} 
& \colhead{$b_2$} &\colhead{profile - \tablenotemark{a}}\\
\colhead{} & \colhead{} & \colhead{(\kms )} & \colhead{} & \colhead{} & \colhead{} & \colhead{} & \colhead{multiplicity\tablenotemark{b}}} 
\startdata
 0.132293 & 0.133064 &            204 &$13.60\pm 0.03$ & $13.35\pm 0.06$ & $ 20.5\pm  1.6$ & $  36.2\pm  6.1 $ & 1 - 3 \\
 0.133064 & 0.133775 &            188 &$13.35\pm 0.06$ & $13.37\pm 0.06$ & $ 36.2\pm  6.1$ & $  47.0\pm  7.8 $ & 1 - 3 \\
 0.152201 & 0.153044 &            219 &$13.52\pm 0.04$ & $13.82\pm 0.03$ & $ 22.5\pm  2.1$ & $  50.8\pm  3.1 $ & 1 - 2 \\
 0.166628 & 0.166962 &  \phantom{1}86 &$13.30\pm 0.11$ & $14.17\pm 0.06$ & $  9.0\pm  2.5$ & $ 112.2\pm  8.3 $ & 2,4 - 3 \\
 0.166628 & 0.167139 &            131 &$13.30\pm 0.11$ & $16.45\pm 0.05$ & $  9.0\pm  2.5$ & $  29.4\pm  1.0 $ & 2,4 - 3 \\
 0.166962 & 0.167139 &  \phantom{1}45 &$14.17\pm 0.06$ & $16.45\pm 0.05$ & $112.2\pm  8.3$ & $  29.4\pm  1.0 $ & 3,4 - 3 \\
 0.245130 & 0.245543 &  \phantom{1}99 &$13.32\pm 0.09$ & $13.72\pm 0.04$ & $ 55.6\pm 14.5$ & $  25.4\pm  2.3 $ & 1,4 - 2	 \\
 0.299040 & 0.299953 &            210 &$13.38\pm 0.07$ & $13.32\pm 0.12$ & $ 53.4\pm  9.7$ & $  72.6\pm 26.4 $ & 1 - 2	 \\
 0.320076 & 0.321160 &            246 &$13.53\pm 0.05$ & $13.30\pm 0.10$ & $ 26.0\pm  3.4$ & $  51.7\pm 15.3 $ & 1 - 2	 \\
 0.341860 & 0.342339 &            107 &$13.59\pm 0.07$ & $13.50\pm 0.09$ & $ 44.4\pm  7.5$ & $  49.8\pm 11.8 $ & 2 - 2	 \\
 0.350988 & 0.351511 &            116 &$14.20\pm 0.03$ & $13.56\pm 0.05$ & $ 40.0\pm  2.4$ & $  27.3\pm   3.8$ & 2,4 - 3	  \\
 0.351511 & 0.352119 &		  135 &$13.56\pm 0.05$ & $13.60\pm 0.05$ & $ 27.3\pm  3.8$ & $  35.1\pm   4.8$ & 1,4 - 3	  \\
 0.360799 & 0.361025 &	 \phantom{1}50 &$15.21\pm 0.02$ & $14.35\pm 0.04$ & $ 22.6\pm  0.6$ & $ 134.1\pm  17.2$ & 3,4 - 2	  \\
 0.399031 & 0.400105 &		  230 &$13.34\pm 0.11$ & $13.30\pm 0.11$ & $ 56.6\pm 18.4$ & $  52.9\pm  17.3$ & 1 - 2	  \\
 0.408875 & 0.409563 &		  146 &$14.36\pm 0.03$ & $13.52\pm 0.08$ & $ 37.3\pm  2.1$ & $  24.6\pm   5.0$ & 1,4 - 2	  \\
\enddata
\vspace{2mm}
\tablenotetext{a}{1 - well-separated; 2 - double-minimum trough  3 - asymmetric profile; 4 - \Lyb\ used}
\tablenotetext{b}{2, 3 ,4 signify pair, triplet, quadruplet}
\end{deluxetable}

\clearpage
\begin{deluxetable}{lcclllcrcl}
\tablewidth{0pc}
\tablecaption{Galaxies around PKS~0405-123
\label{tab:galaxies}}
\tablehead{ 
\colhead{} & \colhead{} & \colhead{} & \multicolumn{2}{c}{mag\tablenotemark{a}} & \colhead{} & \colhead{} & \multicolumn{2}{c}{distance\tablenotemark{d}} & \colhead{} \\
\colhead{number} &
\colhead{$\alpha$ (J2000)} & \colhead{$\delta$ (J2000)} & \colhead{($r$)}  & \colhead{(abs)}     &\colhead{$z$\tablenotemark{b}} 
&\colhead{type\tablenotemark{c}}
&\colhead{arcsec} & \colhead{local Mpc} & \colhead{reference\tablenotemark{e}} \\
} 
\footnotesize
\startdata

 1 &  04:07:27.8 & -12:12:13.6 & 20.5 & -21.6 & 0.4758	& E & 319.9 & 1.900 &		 \\
 2 &  04:07:32.3 & -12:11:07.8 & 21.6 & -19.7 & 0.3410	& E & 249.7 & 1.214 &		 \\
 3 &  04:07:34.5 & -12:10:11.1 & 21.2 & -20.5 & 0.4062	& E & 229.7 & 1.248 &		 \\
 4 &  04:07:35.6 & -12:14:48.5 & 21.7 & -19.5 & 0.3353	& E & 275.5 & 1.323 &		 \\
 5 &  04:07:35.6 & -12:10:12.4 & 21.8 & -20.8 & 0.5745	& E & 213.9 & 1.401 &		 \\
 6 &  04:07:36.3 & -12:11:18.0 & 21.5 & -19.8 & 0.3415	& E & 187.2 & 0.911 &		 \\
 7 &  04:07:36.4 & -12:11:16.3 & 21.9 & -19.4 & 0.3415	& E & 186.0 & 0.905 &		 \\
 8 &  04:07:40.5 & -12:14:20.6 & 20.5 & -19.6 & 0.2089	& E & 205.1 & 0.700 &		 \\
 9 &  04:07:41.0 & -12:13:15.5 & 20.6 & -21.9 & 0.5563	& E & 151.8 & 0.979 &		 \\
10 &  04:07:41.1 & -12:09:13.0 & 21.5 & -20.9 & 0.5235	& E & 181.9 & 1.138 &		 \\
11 &  04:07:41.8 & -12:09:33.9 & 22.5 & -19.0 & 0.3681	& E & 159.5 & 0.815 & [EY94] 105 \\
12 &  04:07:42.5 & -12:15:29.0 & 21.0 & -20.9 & 0.4268	& E & 250.5 & 1.402 &		 \\
13 &  04:07:42.7 & -12:11:32.2 & 19.4 & -20.6 & 0.2038	& E &  88.9 & 0.298 &		 \\
14 &  04:07:42.9 & -12:09:11.6 & 22.1 & -20.7 & 0.6063	& E & 167.7 & 1.126 &		 \\
15 &  04:07:43.9 & -12:10:39.0 & 21.6 & -19.2 & 0.2800	& E &  90.7 & 0.386 & [EY94] 149 \\
16 &  04:07:43.9 & -12:12:08.9 & 18.6 & -20.8 & 0.1580:	& \nodata  &  77.9 & 0.213 & [EY94] 143\tablenotemark{f} \\ 
17 &  04:07:44.3 & -12:11:23.1 & 21.9 & -20.2 & 0.4645	& E &  65.7 & 0.385 &		 \\
18 &  04:07:44.3 & -12:13:22.5 & 22.3 & -20.3 & 0.5730	& E & 124.5 & 0.814 &		 \\
19 &  04:07:44.3 & -12:08:36.1 & 20.6 & -21.7 & 0.5086	& E & 190.7 & 1.176 &		 \\
20 &  04:07:44.4 & -12:12:33.1 & 22.3 & -19.2 & 0.3785	& E &  84.2 & 0.438 &		 \\
21 &  04:07:44.5 & -12:13:52.9 & 20.8 & -22.0 & 0.6235	& E & 150.1 & 1.022 & [EY94] 150 \\
22 &  04:07:45.7 & -12:11:08.9 & 19.0: & -22.4: & 0.3620:	& \nodata  &  49.5 & 0.250 & [EY94] 177\tablenotemark{f} \\ 
23 &  04:07:45.9 & -12:11:00.0 & 21.2 & -21.4 & 0.5696	& A &  53.7 & 0.350 & [EY94] 180 \\
24 &  04:07:46.2 & -12:15:56.3 & 21.1 & -21.5 & 0.5725	& A & 262.5 & 1.716 &		 \\
25 &  04:07:46.7 & -12:14:02.0 & 20.9 & -20.8 & 0.4065	& A & 148.6 & 0.808 & [EY94] 195 \\
26 &  04:07:46.7 & -12:10:07.0 & 22.5 & -20.4 & 0.6410	& E &  93.2 & 0.642 & [EY94] 200 \\
27 &  04:07:46.7 & -12:12:16.0 & 20.9 & -20.5 & 0.3514	& E &  48.4 & 0.240 &		 \\
28 &  04:07:46.9 & -12:10:28.0 & 22.3 & -20.7 & 0.6552	& E &  72.3 & 0.503 & [EY94] 207 \\
29 &  04:07:47.8 & -12:13:47.9 & 20.9 & -20.8 & 0.3968	& E & 132.4 & 0.709 & [EY94] 219 \\
30 &  04:07:47.8 & -12:13:46.7 & 21.1 & -22.0 & 0.6885	& E & 131.2 & 0.932 &		 \\
31 &  04:07:48.0 & -12:06:27.8 & 22.0 & -13.0 & 0.0234:	& E & 308.3 & 0.144 &		 \\
32 &  04:07:48.2 & -12:11:48.9 & 20.0 & -22.6 & 0.5696	& A &  13.8 & 0.090 & [EY94] 231 \\
33 &  04:07:48.3 & -12:11:02.0 & 21.0 & -18.5 & 0.1670:	& E &  34.1 & 0.097 & [EY94] 241\tablenotemark{f} \\ 
34 &  04:07:48.4 & -12:12:10.9 & 21.5 & -19.9 & 0.3520	& A &  35.0 & 0.174 & [EY94] 240 \\
35 &  04:07:48.6 & -12:11:51.0 & 21.5 & -21.1 & 0.5657	& A &  15.1 & 0.098 & [EY94] 244 \\
36 &  04:07:48.8 & -12:11:32.0 & 21.6 & -21.0 & 0.5709	& E &	7.3 & 0.048 & [EY94] 249 \\
37 &  04:07:49.1 & -12:11:43.0 & 21.1 & -21.5 & 0.5714	& E &  11.6 & 0.076 & [EY94] 258 \\
38 &  04:07:49.1 & -12:12:02.9 & 21.6 & -21.0 & 0.5779	& A &  28.5 & 0.187 & [EY94] 253 \\
39 &  04:07:49.4 & -12:12:10.9 & 21.9 & -20.7 & 0.5777	& A &  37.6 & 0.247 & [EY94] 266 \\
40 &  04:07:49.5 & -12:12:21.0 & 22.0 & -20.8 & 0.6167	& E &  47.5 & 0.322 & [EY94] 269 \\
41 &  04:07:49.6 & -12:12:45.9 & 21.6 & -20.7 & 0.5170	& E &  72.4 & 0.450 & [EY94] 274 \\
42 &  04:07:50.0 & -12:09:50.9 & 19.8 & -21.4 & 0.3255	& A & 107.8 & 0.507 & [EY94] 299 \\
43 &  04:07:50.6 & -12:12:24.0 & 19.7 & -21.2 & 0.2973	& E &  57.8 & 0.256 & [EY94] 300 \\
44 &  04:07:50.9 & -12:15:48.7 & 21.4 & -21.7 & 0.6831	& E & 255.5 & 1.810 &		 \\
45 &  04:07:51.2 & -12:11:37.0 & 17.8:& -21.7:& 0.1667:	& A &  41.4 & 0.118 & [EY94] 309\tablenotemark{f}\tablenotemark{g} \\ 
46 &  04:07:51.2 & -12:16:31.2 & 20.7 & -21.0 & 0.4063	& A & 298.4 & 1.621 &		 \\
47 &  04:07:51.5 & -12:12:55.9 & 21.1 & -20.7 & 0.4248	& E &  92.3 & 0.515 & [EY94] 311 \\
48 &  04:07:51.6 & -12:08:20.7 & 22.0 & -17.0 & 0.1342	& E & 201.1 & 0.480 &		 \\
49 &  04:07:51.7 & -12:08:41.2 & 21.5 & -20.9 & 0.5239	& E & 181.6 & 1.137 &		 \\
50 &  04:07:51.8 & -12:13:14.9 & 19.8 & -22.0 & 0.4242	& E & 111.2 & 0.620 & [EY94] 319 \\
51 &  04:07:51.8 & -12:13:16.9 & 20.6 & -21.2 & 0.4241	& E & 113.0 & 0.630 & [EY94] 318 \\
52 &  04:07:52.3 & -12:14:07.0 & 19.9 & -21.1 & 0.3085	& A & 161.9 & 0.734 & [EY94] 326 \\
53 &  04:07:52.4 & -12:08:08.5 & 22.0 & -20.9 & 0.6424	& E & 216.4 & 1.493 &		 \\
54 &  04:07:52.5 & -12:15:48.1 & 20.5 & -17.6 & 0.0913	& E & 259.8 & 0.441 &		 \\
55 &  04:07:52.8 & -12:06:46.4 & 19.6 & -20.9 & 0.2461	& E & 297.2 & 1.150 &		 \\
56 &  04:07:53.6 & -12:12:28.2 & 21.9 & -20.9 & 0.6080	& E &  94.8 & 0.637 &		 \\
57 &  04:07:53.8 & -12:13:56.9 & 21.3 & -20.1 & 0.3518	& E & 162.8 & 0.808 & [EY94] 356 \\
58 &  04:07:53.8 & -12:15:59.1 & 21.6 & -20.2 & 0.4252	& E & 275.8 & 1.540 &		 \\
59 &  04:07:54.2 & -12:08:54.0 & 22.0 & -19.1 & 0.3200	& E & 184.1 & 0.856 & [EY94] 376 \\
60 &  04:07:54.2 & -12:14:43.2 & 19.6 & -18.6 & 0.0964	& E & 206.6 & 0.369 &		 \\
61 &  04:07:54.8 & -12:12:10.5 & 22.1 & -20.8 & 0.6413	& E & 103.0 & 0.710 &		 \\
62 &  04:07:55.0 & -12:09:10.9 & 20.5 & -22.1 & 0.5604	& E & 176.0 & 1.139 & [EY94] 391 \\
63 &  04:07:55.0 & -12:11:42.6 & 21.9 & -21.0 & 0.6405	& A & 100.7 & 0.694 &		 \\
64 &  04:07:55.5 & -12:10:37.0 & 20.5 & -21.4 & 0.4282	& E & 122.6 & 0.687 & [EY94] 398 \\
65 &  04:07:55.8 & -12:14:41.1 & 20.3 & -18.6 & 0.1312	& E & 216.7 & 0.507 &		 \\
66 &  04:07:55.9 & -12:13:37.9 & 22.1 & -21.0 & 0.6863	& A & 167.7 & 1.190 &		 \\
67 &  04:07:57.2 & -12:14:40.4 & 21.3 & -20.1 & 0.3610	& E & 228.2 & 1.151 &		 \\
68 &  04:07:57.9 & -12:14:21.2 & 22.1 & -20.2 & 0.5027	& E & 219.5 & 1.346 &		 \\
69 &  04:07:59.5 & -12:10:31.6 & 20.6 & -21.7 & 0.5092	& A & 181.7 & 1.121 &		 \\
70 &  04:08:01.7 & -12:10:15.2 & 22.9 & -17.9 & 0.2797	& E & 218.6 & 0.929 &		 \\
71 &  04:08:04.0 & -12:12:28.9 & 20.9 & -22.0 & 0.6387	& A & 244.4 & 1.682 &		 \\
72 &  04:08:05.4 & -12:13:04.9 & 21.5 & -16.3 & 0.0791	& E & 275.1 & 0.409 &		 \\
73 &  04:08:10.2 & -12:12:04.8 & 23.4 & -20.0 & 0.7751	& E & 334.4 & 2.486 &		 \\

\enddata
\vspace{2mm}
\tablenotetext{a}{Errors are $\sim 0.1$ magnitudes. Absolute magnitudes calculated using distance modulus from our adopted cosmology.}
\tablenotetext{b}{Errors are $\sim 100$~\kms\ in the rest frame.}
\tablenotetext{c}{Types are: A - absorption line spectrum; E - emission line spectrum.}
\tablenotetext{d}{Distance from PKS~0405-123 in arcsec and rest frame Mpc.}
\tablenotetext{e}{\phantom{}Galaxies are from this work unless otherwise
noted.  Listings are from the NASA Extragalactic Database (NED).  
[EY94] - \cite{Ellingson94}}
\tablenotetext{f}{\phantom{}Listed in both \citet{Spinrad93} and \citet{Ellingson94}.  Redshifts are
from \citeauthor{Spinrad93} and are relatively less certain.  Photometry is consistent between
the two papers except for galaxy 
177, where $\Delta r = 0.1$~mag.  We take the \citeauthor{Ellingson94} value for consistency.
The EY93 nomenclature in NED is an error.}
\tablenotetext{g}{Blended with star.  Take magnitude as lower limit for galaxy. Blended nature and galaxy type
from J. Prochaska (private communication, 2005).}
\end{deluxetable}

\clearpage
\begin{deluxetable}{lcccccl}
\tablewidth{0pc}
\tablecaption{AGN or galaxies near \Lya\ forest voids
\label{tab:AGN}}
\tablehead{ 
\colhead{name\tablenotemark{a}} & \colhead{RA, dec (J2000)}     &\colhead{$z$} & \colhead{arcmin\tablenotemark{b}} 
& \colhead{Mpc\tablenotemark{c}} & \colhead{type} & \colhead{reference\tablenotemark{d}}} 
\footnotesize
\startdata
2MASX J04064568-1244205  &      04:06:45.5 -12:44:22& 0.0326   &   36.2  & 1.41  & galaxy & 1\\
2MASX J04100991-1124129  &      04:10:09.8 -11:24:13& 0.0333   &   58.7  & 2.35  & galaxy & 1\\
NPM1G-13.0163            &      04:08:31.2 -12:59:15& 0.0789   &   48.8  & 4.39  & Sy1.9  & 1, 2 \\
2MASX J04053763-1259048  &      04:05:37.6 -12:59:05& 0.1185   &   57.2  & 7.38  & galaxy & 1, 3 \\
1WGA J0405.5-1258	 &      04:05:34.9 -12:58:40& 0.1205   &   57.3  & 7.45  & AGN    & 4 \\
\enddata
\vspace{2mm}
\tablenotetext{a}{\phantom{.}$\!$Name as used in \citet{Veron03} or NED.}
\tablenotetext{b}{Distance from PKS~0405-123 in arcmin.}
\tablenotetext{c}{Distance from PKS~0405-123 in local frame Mpc.}
\tablenotetext{d}{1-\citet{Jarrett04}; 2-\citet{Moran96}; 3-\citet{Stoll93}; 4-\citet{delOlmo91}}
\end{deluxetable}

\clearpage


\clearpage
\begin{figure}
\plotone{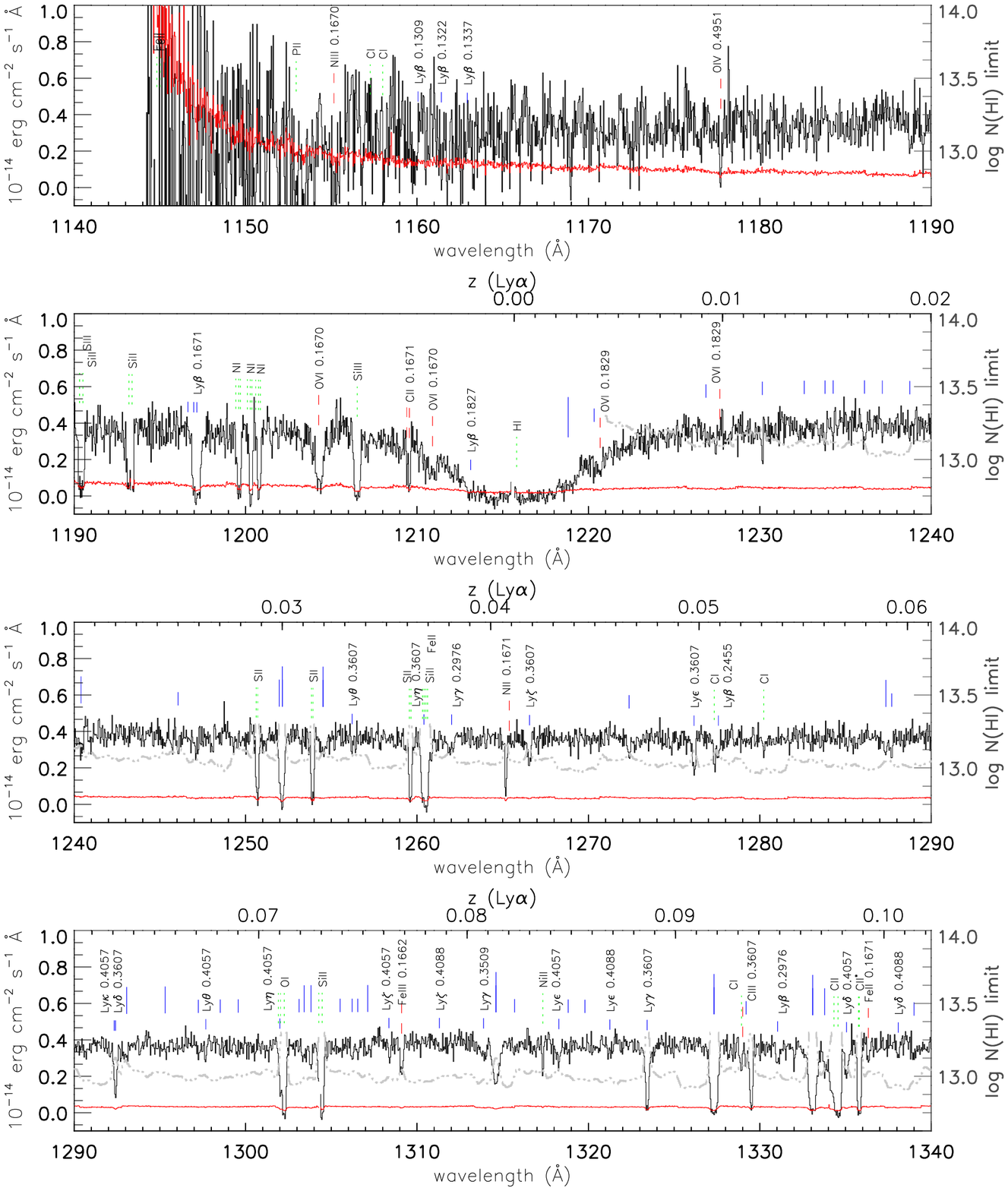}
\vspace*{-5mm}
\caption[PKS~0405--123 STIS spectrum]{PKS~0405--123 
data and $1\sigma$ errors {\it (red)} binned
by 3 pixels for presentation only. 
{\it Grey dash-triple dotted curve:} 80\% detection threshold in $\log N(\HI)$
(right axis).
{\it Long, bold (red) ticks:} \Lya\ lines with metals. 
{\it Long/medium/short unlabelled (blue) ticks:} \Lya\ forest ($z$ on upper axis)
with $\log \NHI\geq 13.3$, $13.1\leq \log \NHI< 13.3$ in the weak
survey (\S~\ref{sec:observations}), other \Lya\ lines respectively.
Higher order Lyman lines are labelled.
{\it Medium dashed (red) ticks:} intervening metal lines.
{\it Medium dotted (green) ticks:} Galactic metal absorption.
\label{fig:plotspec}
}
\end{figure}

\clearpage
\setcounter{figure}{0}
\begin{figure}
\plotone{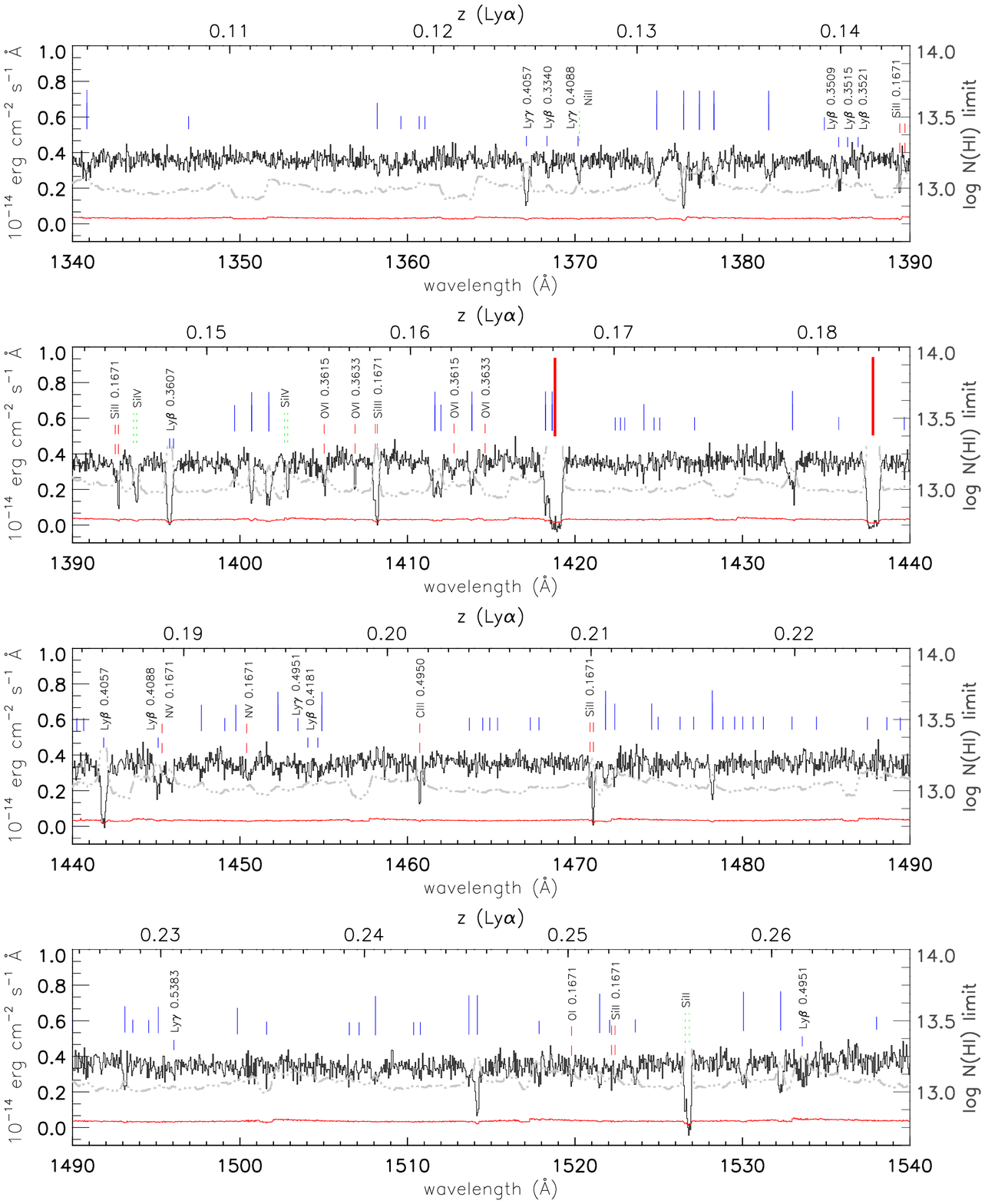}
\caption[PKS~0405--123 STIS spectrum]{
Continued.
}
\end{figure}

\setcounter{figure}{0}
\clearpage
\begin{figure}
\plotone{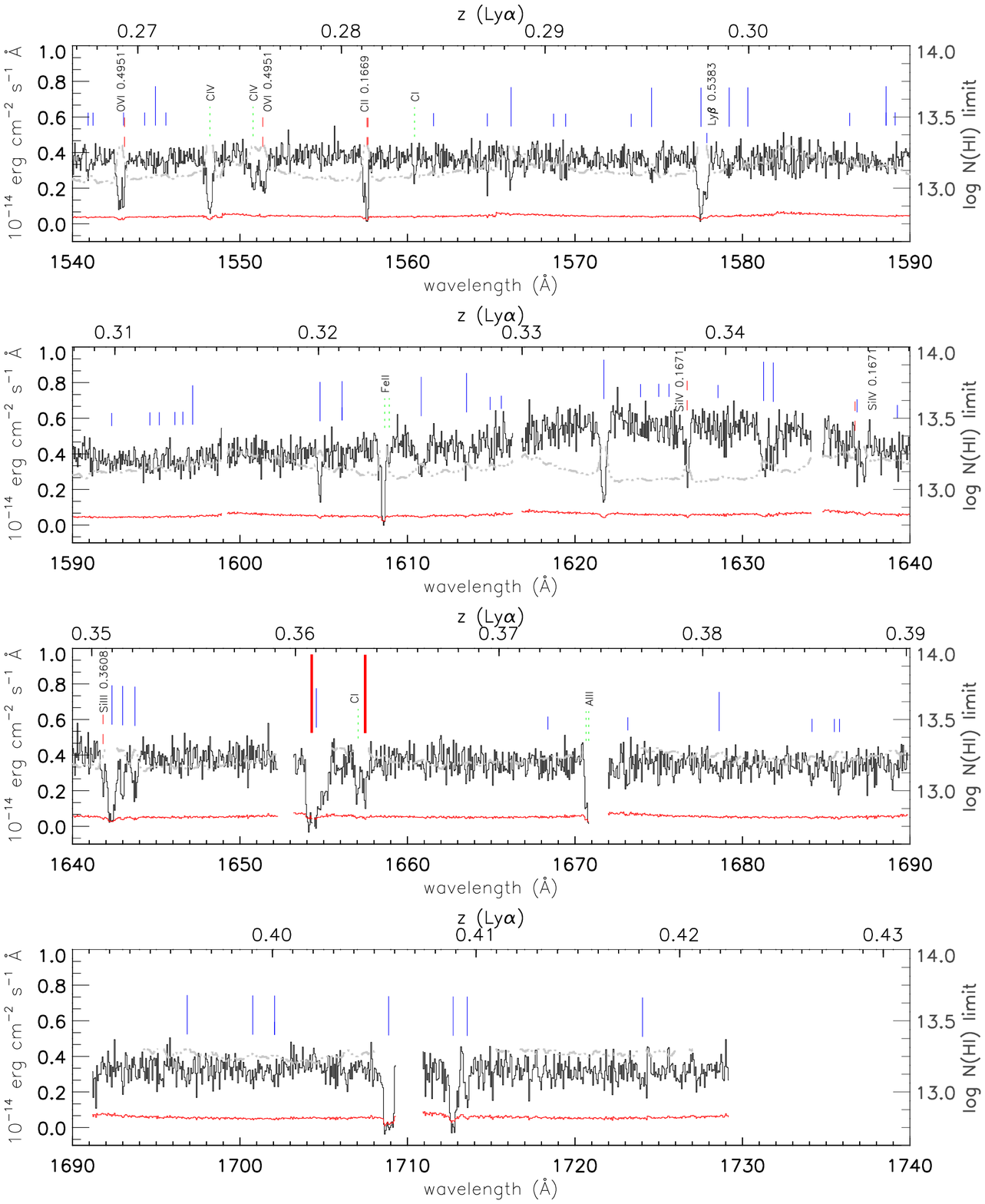}
\caption[PKS~0405--123 STIS spectrum]{
Continued.
}
\end{figure}

\clearpage
\begin{figure}
\epsscale{1.0}
\plotone{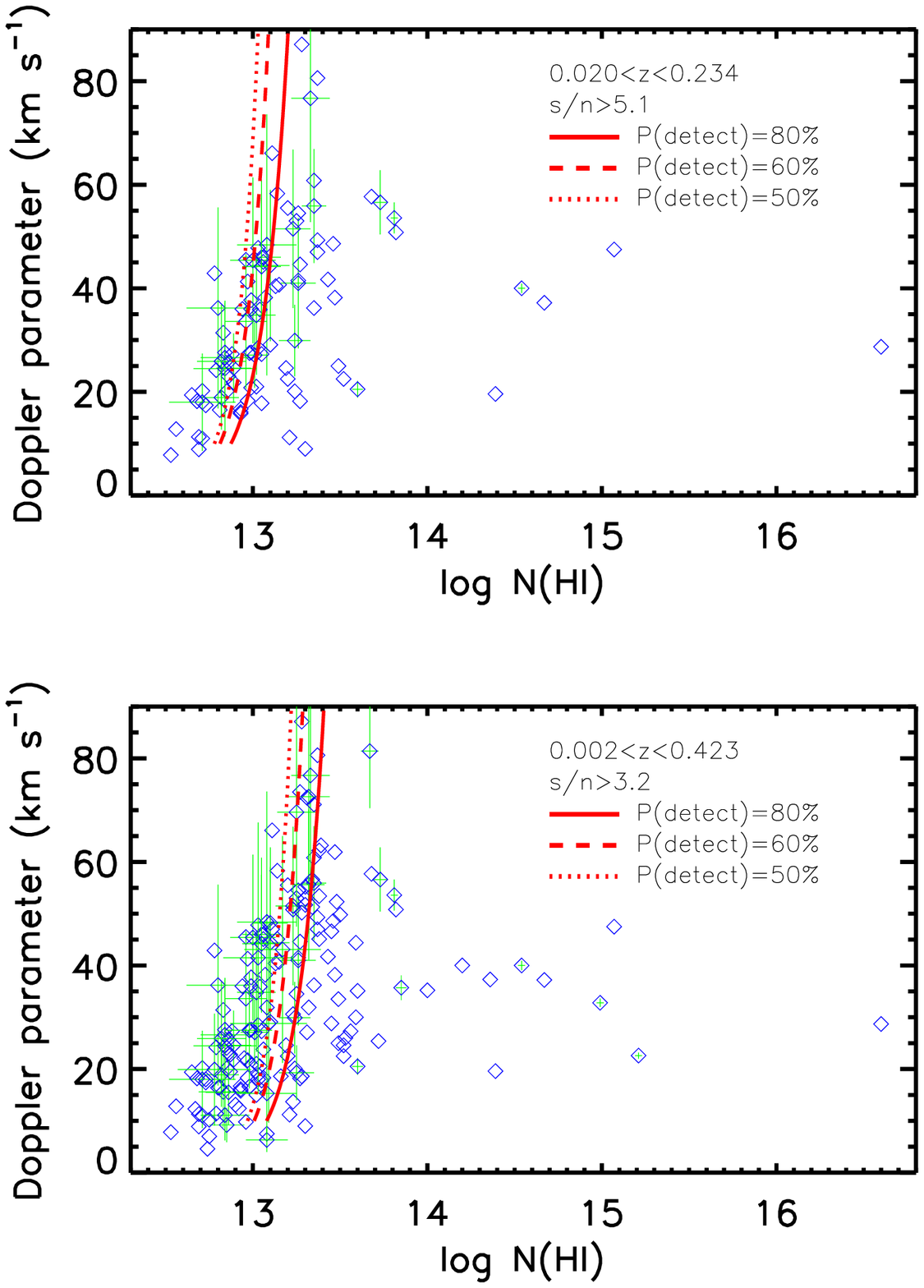}
\caption{\HI\ column densities and Doppler parameters.
{\it Top:} 
\Lya\ forest sample over $0.020<z<0.234$, which is the redshift range
used for
our sample of $\log \NHI\geq 13.1$ (s/n ratio $>5.1$ per pixel, $b=40$~\kms ).  
{\it Open diamonds:} data points.
For clarity, only error bars for every third point are
shown.  {\it Solid, dashed and dotted lines} indicate the boundaries of the
80\%, 60\% and 50\% detection probabilities from 
simulations in \S~\ref{sec:simulatedspectra}.
{\it Below:} As above, but for the 
\Lya\ forest sample over $0.002<z<0.423$, which is used for
our sample of $\log \NHI\geq 13.3$ (s/n ratio $>3.2$ per pixel, $b=40$~\kms ).
\label{fig:nb}
}
\end{figure}

\clearpage
\begin{figure}
\epsscale{1.0}
\plotone{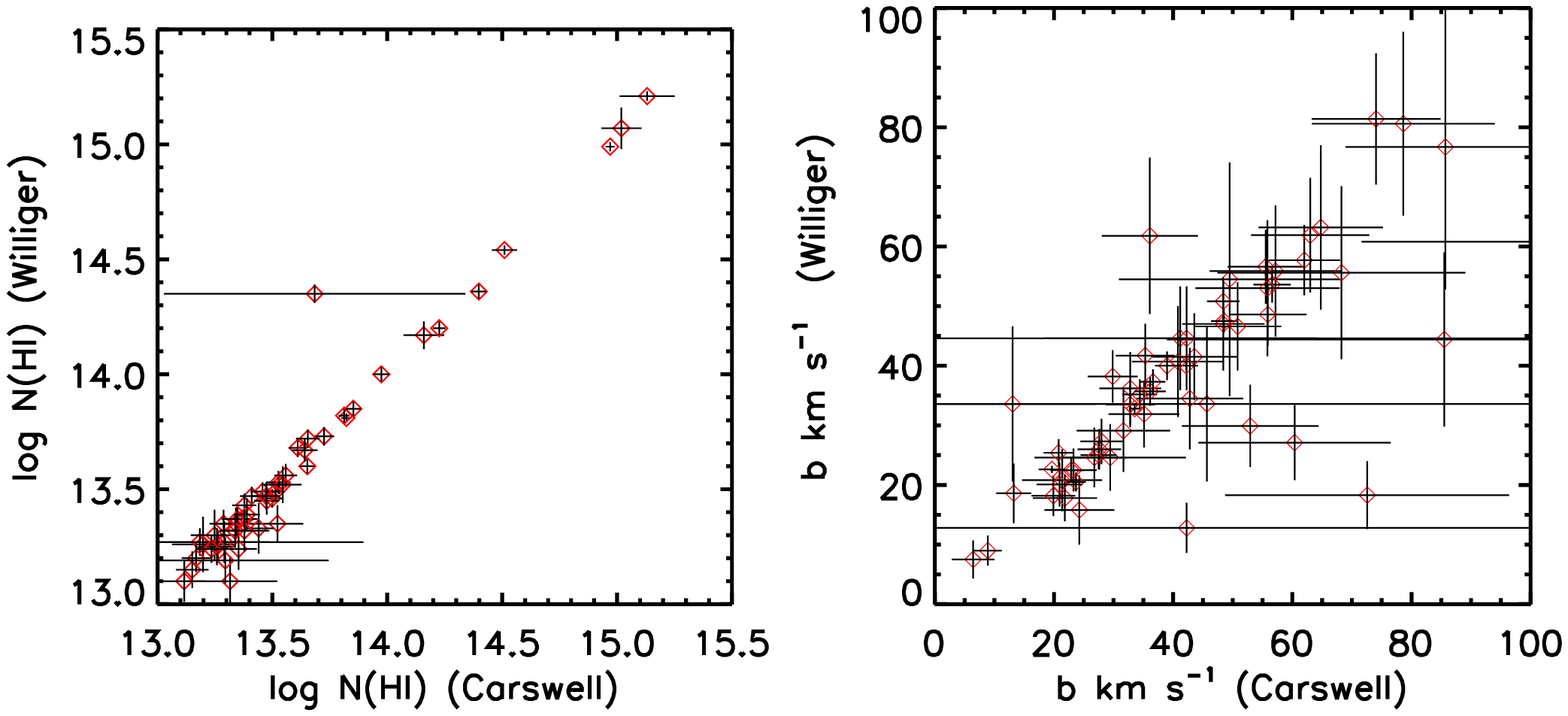}
\caption{Comparison of \HI\ column densities and Doppler parameters.
{\it Left:} \HI\ column densities for the fits done by
Carswell and Williger.  {\it Right:} Doppler parameters.
Only systems in the strong and weak samples, and
for which similar sets of Lyman series lines were fitted, 
are shown.
\label{fig:compare_rfc_gmw}
}
\end{figure}

\clearpage
\begin{figure}
\plotone{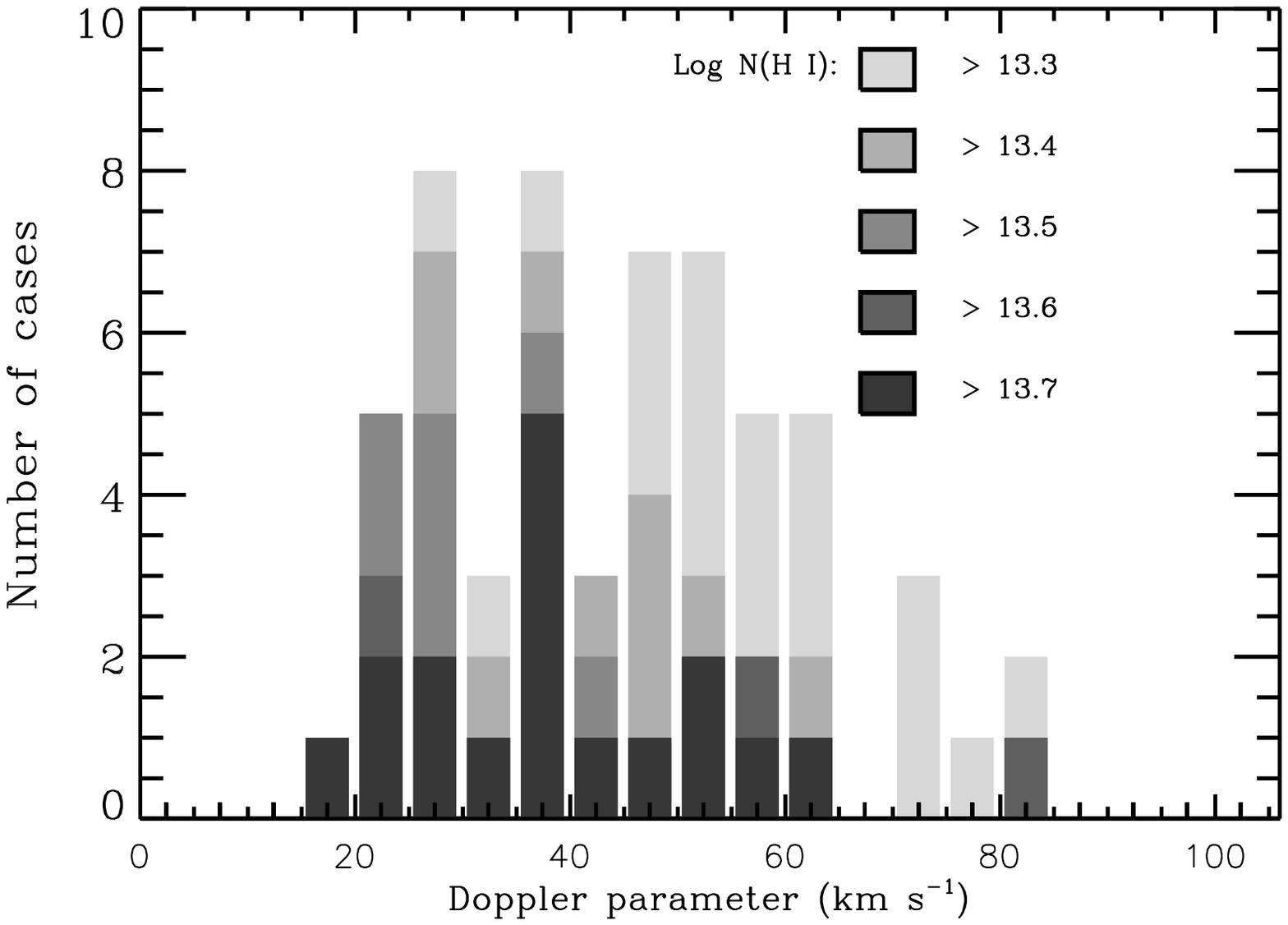}
\caption{Ly$\alpha$ Doppler parameter distribution for a series of
minimum 
$\log \NHI$ thresholds.
The tail at high values is likely a result of unresolved blends.  
\label{fig:dndb}
}
\end{figure}

\clearpage
\begin{figure}
\plotone{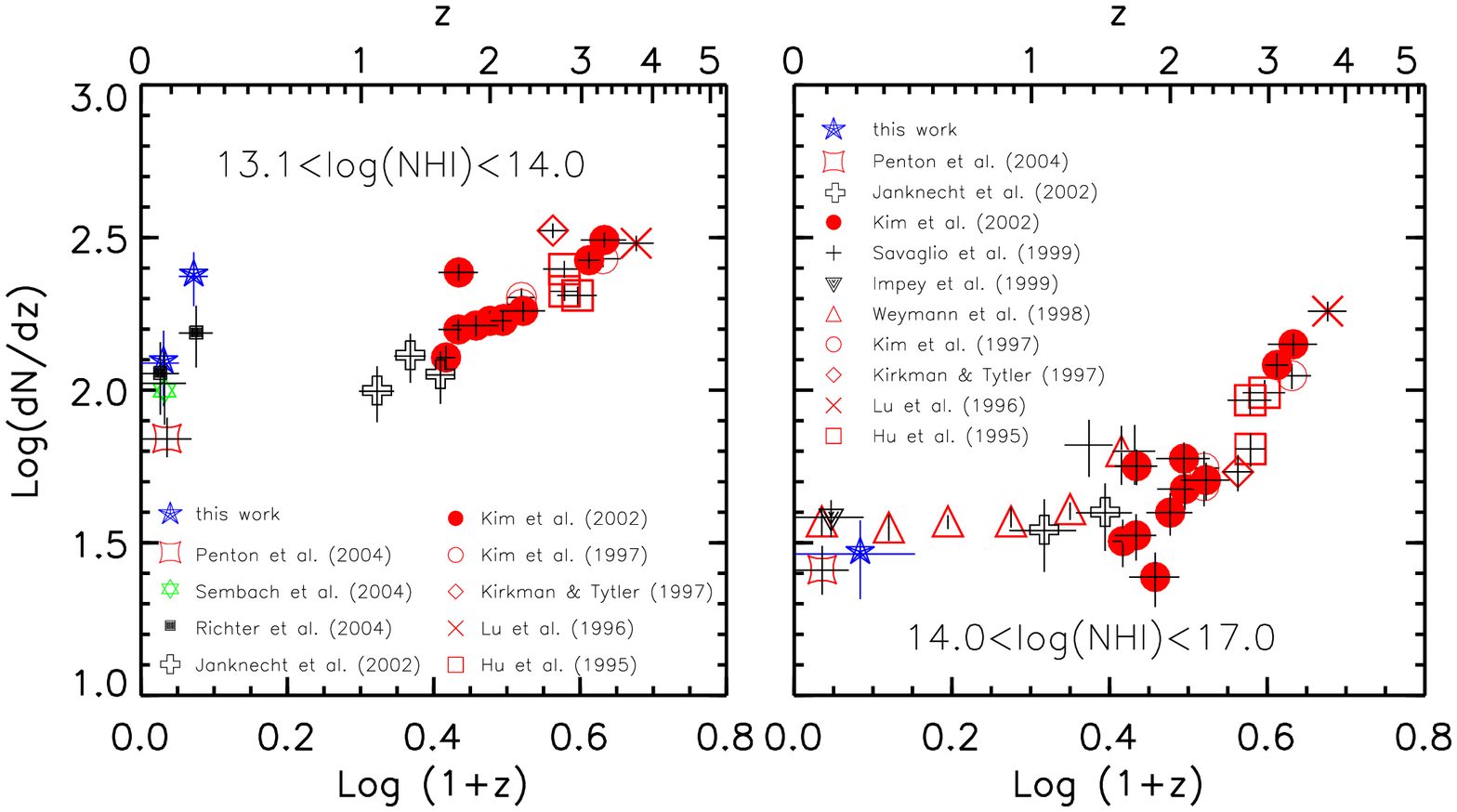}
\caption{{\it Left:} Ly$\alpha$ forest redshift density $dN/dz$ for $13.1<\log \NHI<14.0$.
The high and low redshift stars correspond to
high and low redshift halves of our sample, respectively, 
and show clear evidence of cosmic variance.
{\it Right:} 
$14.0<\log \NHI<17.0$ (right).  Error bars show redshift bins and Poissonian errors in $dN/dz$.
\label{fig:plotkim}
}
\end{figure}

\clearpage
\begin{figure}
\plotone{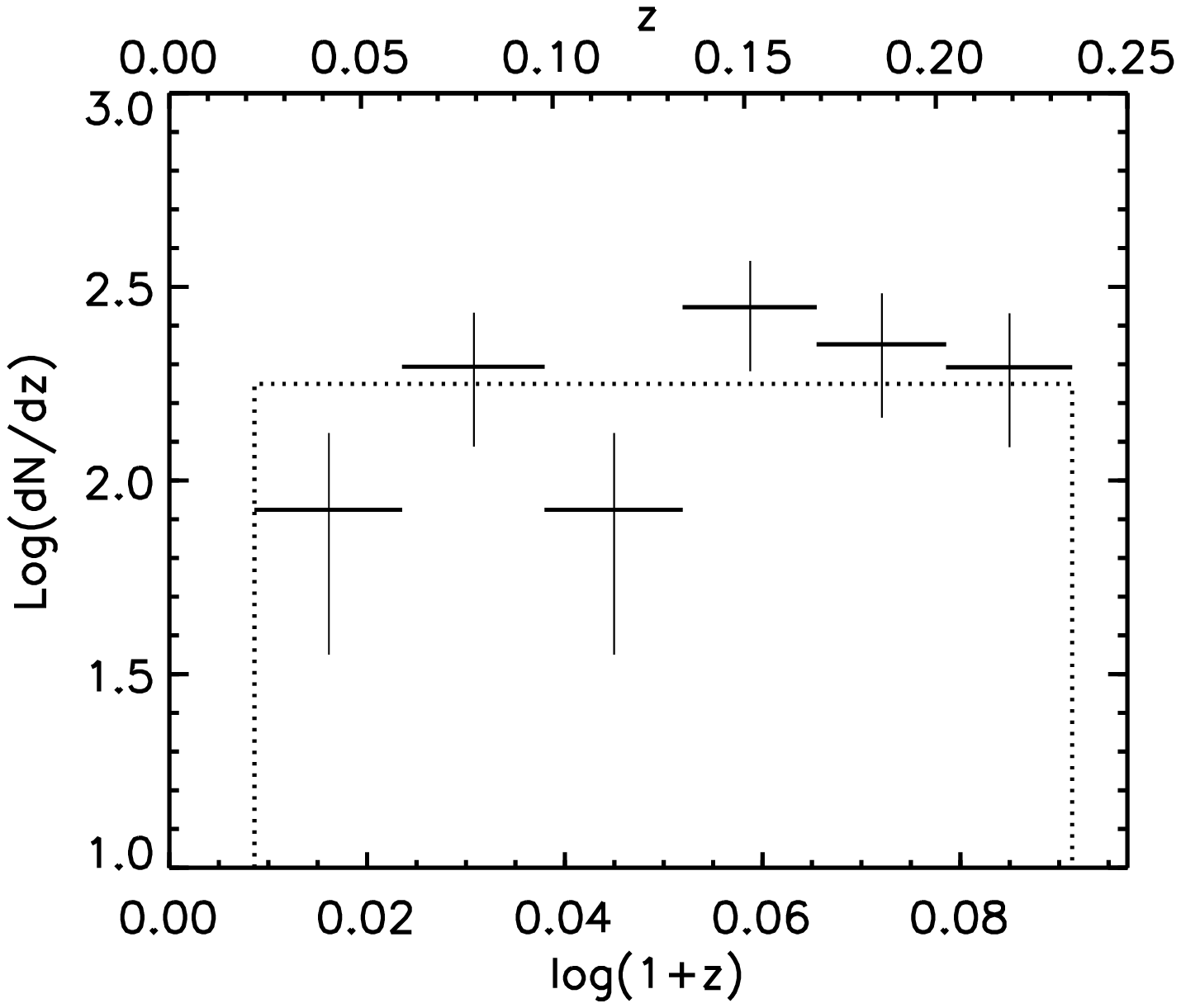}
\caption{Redshift density for \Lya\ absorbers with $13.1\leq \log \NHI< 14.0$ where the
detection probability for a system with $\log \NHI= 13.1$ is 80\% .  Errors are based on
Poisson statistics.  {\it Dotted line:}
expected value for uniform distribution in $z$.
\label{fig:plotdndz_weak}
}
\end{figure}

\clearpage
\begin{figure}
\plotone{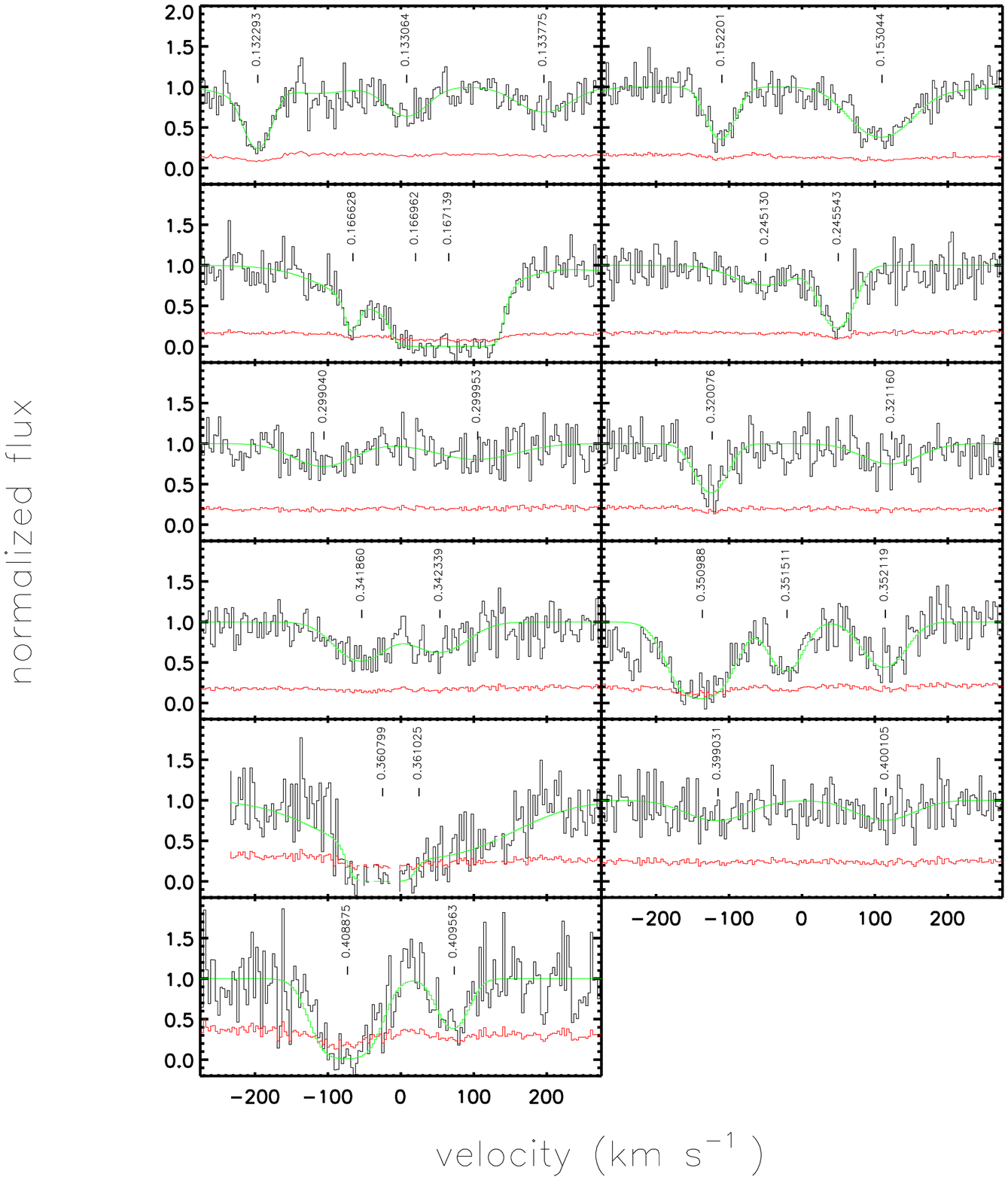}
\caption{\Lya\ pairs with velocity separation $\Delta v<250$~\kms\ for the strong sample.
The data, $1\sigma$ error array and Voigt profile fit are shown.  \Lya\ lines are marked
by ticks and redshifts.  For clarity, 
noise spikes have been suppressed.  
\label{fig:plot_lyapairs}
}
\end{figure}

\clearpage
\begin{figure}
\plotone{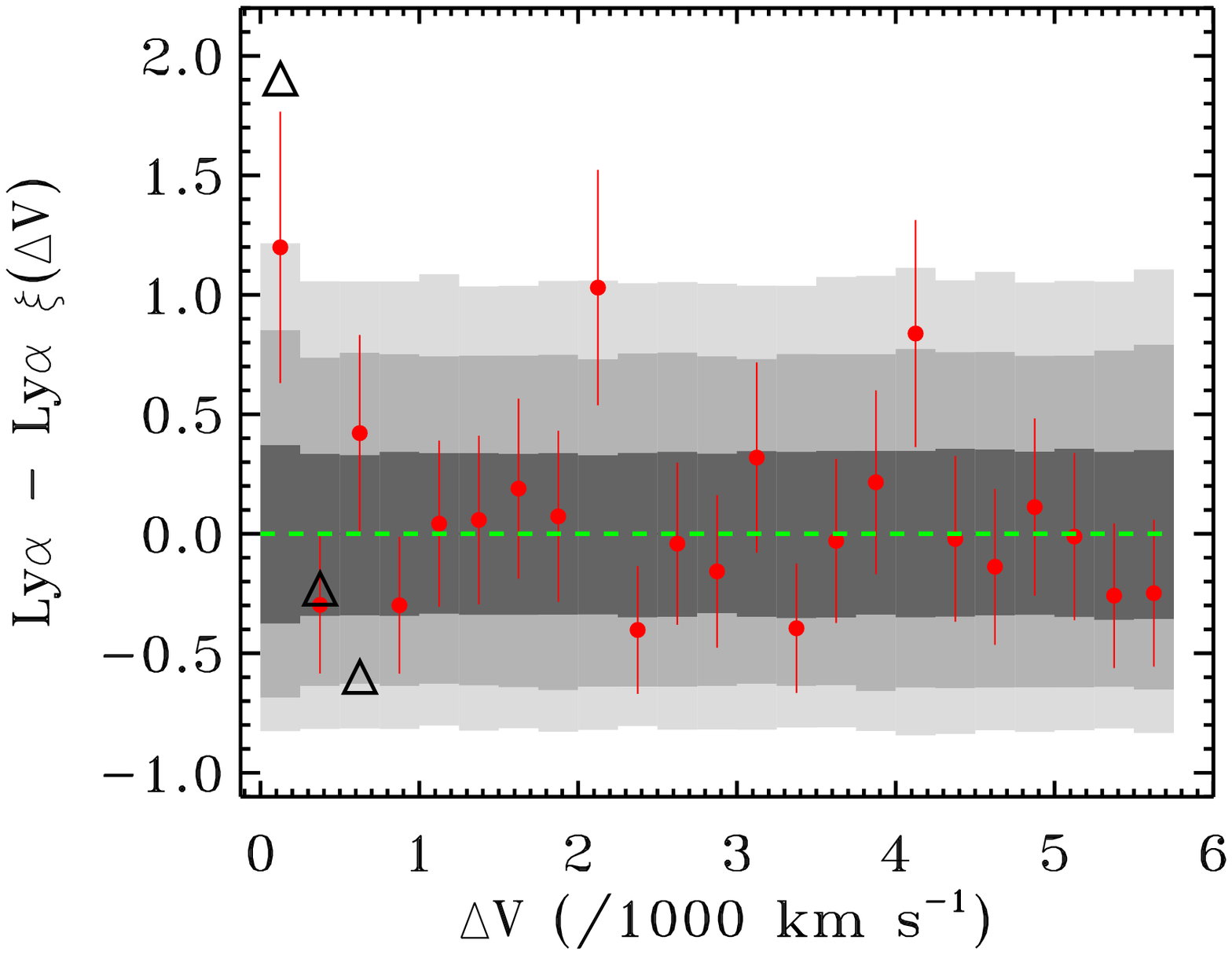}
\caption{Two point correlation function for the strong \Lya\ sample ($\log \NHI\geq 13.3$), which is
the column density threshold with
the most significant signal.
Shaded regions denote 68, 95, 99\% confidence limits from 10$^4$ Monte Carlo simulations.
The value at $\Delta v < 250$~\kms\ has a probability of occurrence from unclustered data of
$P=0.005$.
Simulated line lists have been filtered to eliminate pairs with velocity splittings of $\Delta v < 43$~\kms .
Error bars show example $1\sigma$ Poissonian errors.
The triangles show expected values
from the simulations of \citet{Dave03} for the same column density threshold;
the simulation box artificially
lowers the value of $\xi$ in the higher two velocity bins.  The data differ from the simulation at $\Delta v < 250$~\kms\
at the 1.2~$\sigma$ level.
\label{fig:twoptcorr}
}
\end{figure}

\clearpage
\epsscale{1.1}
\begin{figure}
\caption{CFHT r-band image centered on PKS~0405--123.  The field 
size is 9~arcmin (2.9 local frame Mpc at $z=0.4$), 
and north and east are indicated by the compass.  Galaxies
correspond to numbers in Table~\ref{tab:galaxies}.  PKS~0405--123 is designated by Q.
\label{fig:galaxies}
}
\end{figure}

\clearpage
\begin{figure}
\epsscale{0.5}
\includegraphics[scale=1.0,angle=-90]{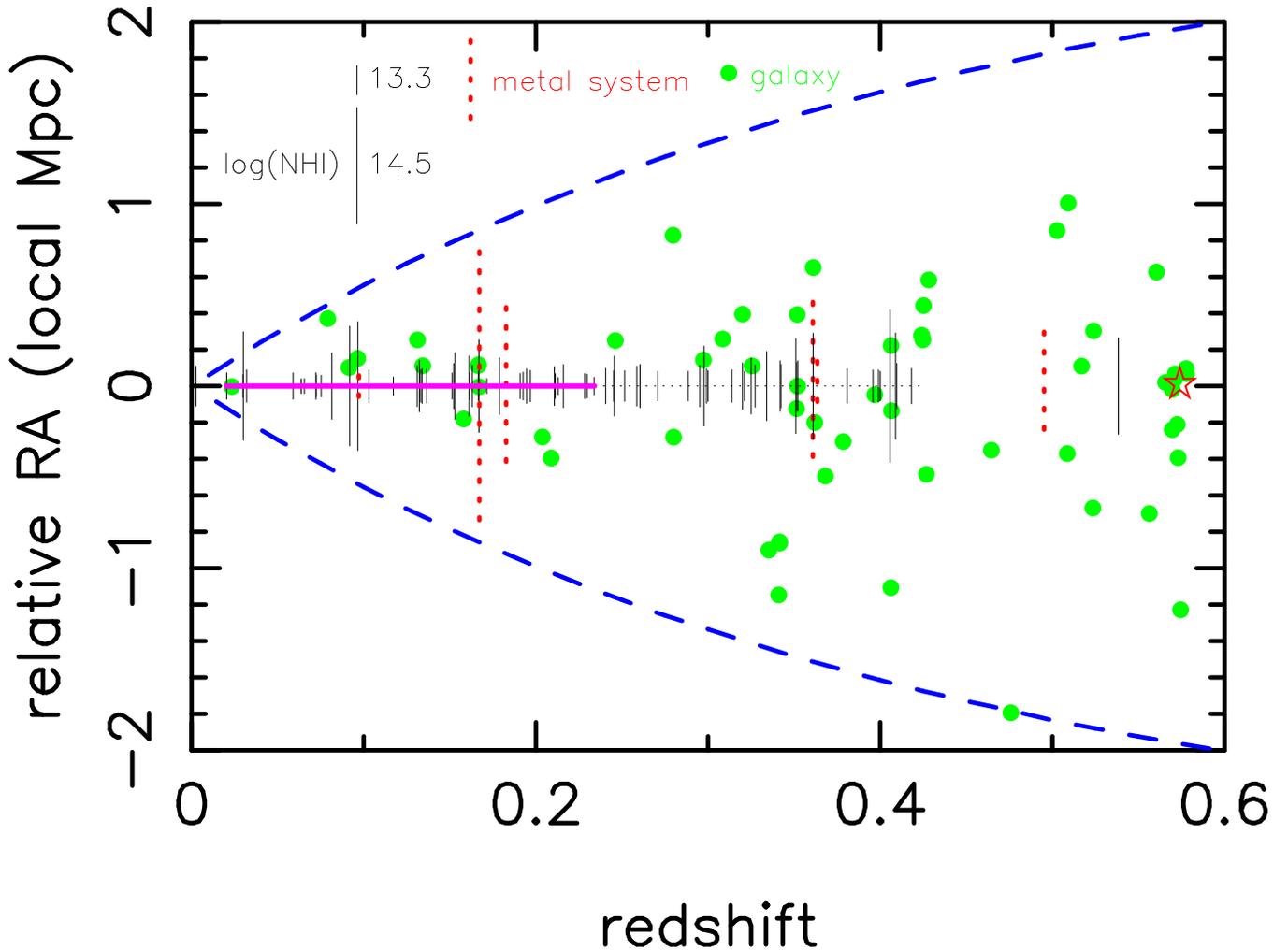}
\caption{Ly$\alpha$ absorbers and galaxies toward PKS~0405--123, projected 
onto the declination plane.
{\it Vertical lines:}
Ly$\alpha$
systems, with length proportional to $\log \NHI$.
{\it Dashed lines and arrows:} metal absorbers.
{\it Dots:} galaxies.  {\it Thick line at RA=0:} $z$ limits where
survey is complete to $\log \NHI=13.1$; only absorbers corresponding 
to the
local 80\% detection probability threshold are shown.  {\it Thin line at RA=0:} $z$ limits 
where survey is complete to $\log \NHI=13.3$.
{\it Star:} PKS~0405--123.  {\it Dashed cone:} 
scope corresponding to a 5~arcmin radius.  
\label{fig:galconera}
}
\end{figure}

\clearpage
\begin{figure}
\epsscale{0.5}
\includegraphics[scale=1.0,angle=-90]{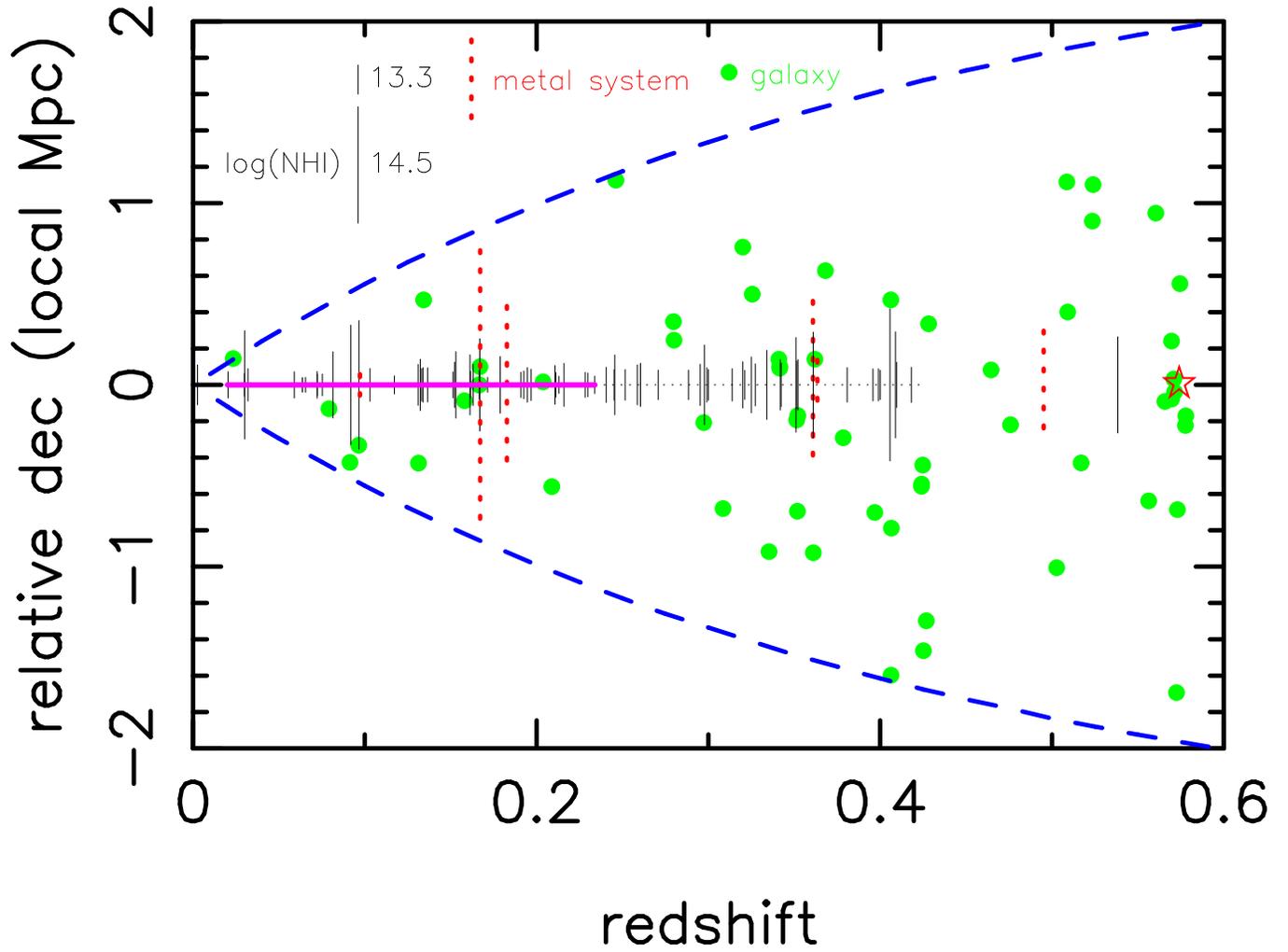}
\caption{As Fig.~\ref{fig:galconera}, but projected onto the right ascension plane.
\label{fig:galconedec}
}
\end{figure}

\clearpage
\begin{figure}
\epsscale{1.0}
\plotone{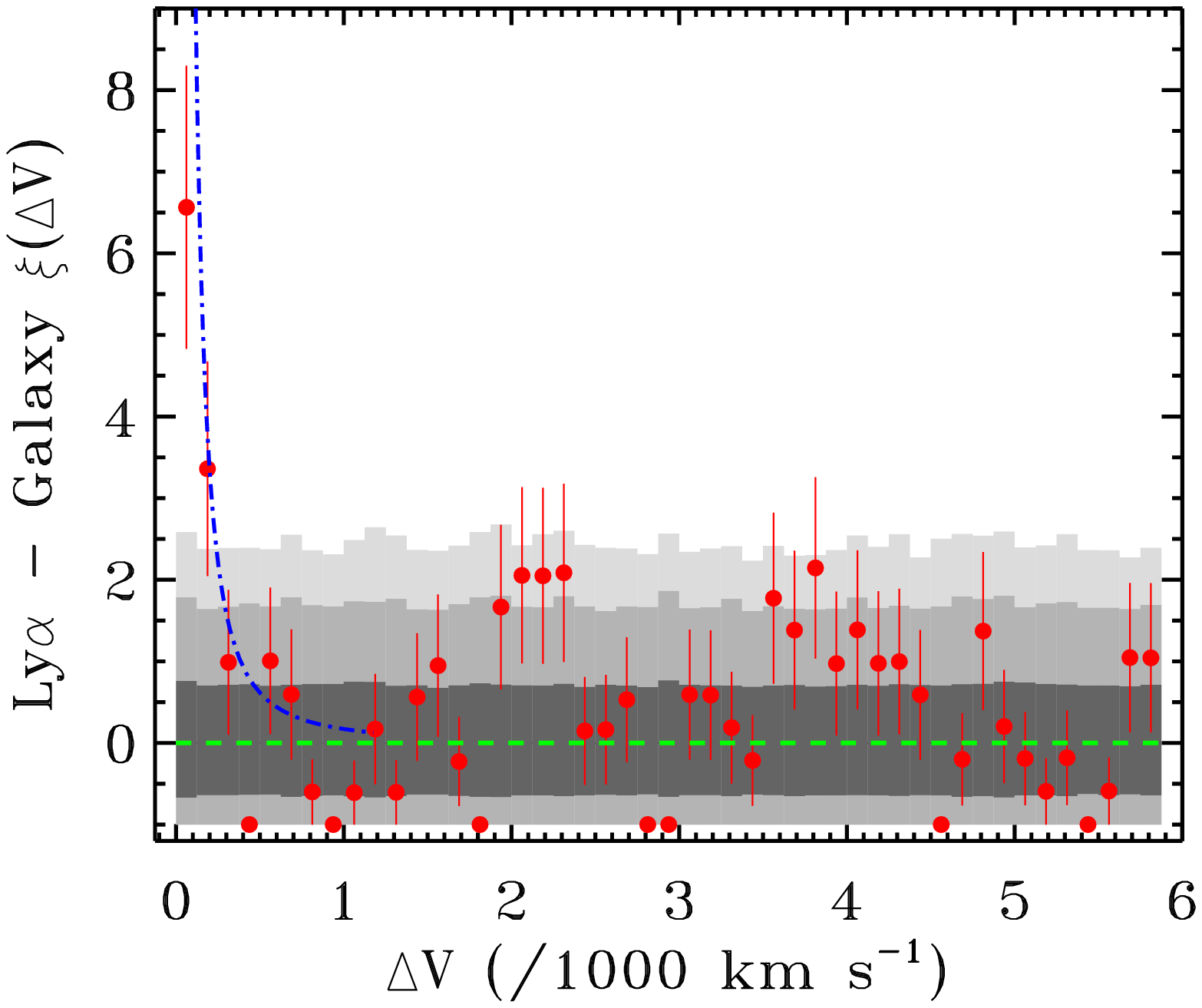} 
\caption{Two point
correlation function in velocity space for the Ly$\alpha$ forest and galaxies
within 2 Mpc projected distance, with $\log \NHI\geq 13.5$, the threshold at
which the correlation significance is maximum. {\it Shaded regions:} 68, 95,
99\% confidence limits from $10^4$ Monte Carlo simulations. {\it Dash-dotted 
line:}
Galaxy-galaxy correlation function from Table~1 of \cite{Zehavi04}, weighted by
the number of galaxies, interpolated for $M_r=-20$ and set to $z=0.2$. Error
bars show example $1\sigma$ Poissonian errors. 
\label{fig:lyagalcorr}
}
\end{figure}

\clearpage
\begin{figure}
\epsscale{1.0}
\plotone{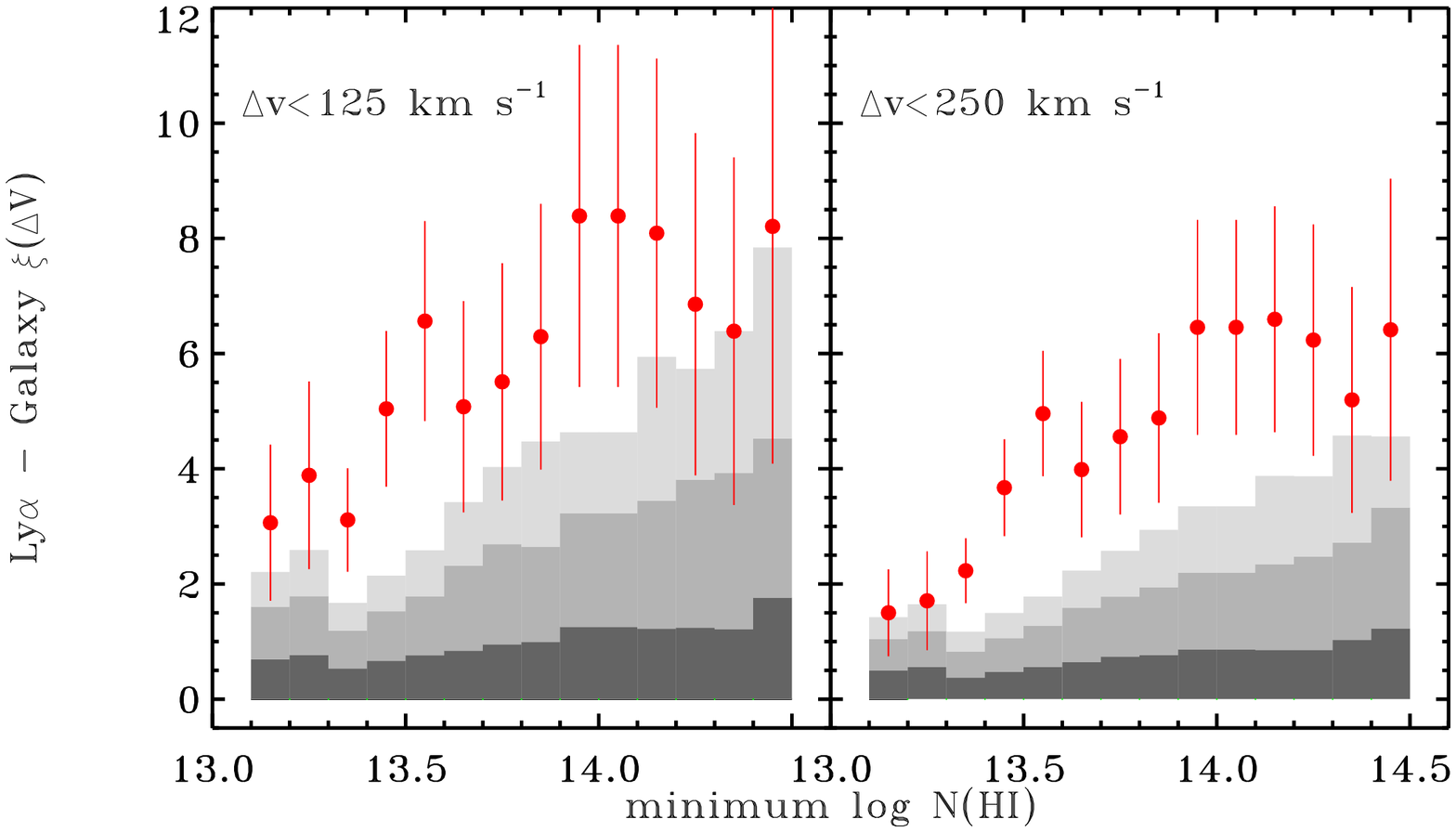} 
\caption{Two point correlation function in velocity space for the Ly$\alpha$ forest and
galaxies within 2 Mpc projected distance, as a function of minimum absorber \HI\ column
density threshold, for $\Delta v < 125$~\kms\ (left panel) and $\Delta v < 250$~\kms\
(right panel).  Symbols and shading  are as in Figure~\ref{fig:lyagalcorr}. 
There is a marginal effect that the weakest
absorbers have their strongest signal at $\Delta v < 125$~\kms , while the strongest ones
strengthen the significance of their signal out to $\Delta v < 250$~\kms .  Such behavior
would be consistent with higher column density systems having a longer correlation length
with galaxies, which would be expected from larger density perturbations.
\label{fig:lyagalcorr_grid}
}
\end{figure}

\clearpage
\begin{figure}
\epsscale{1.0}
\plotone{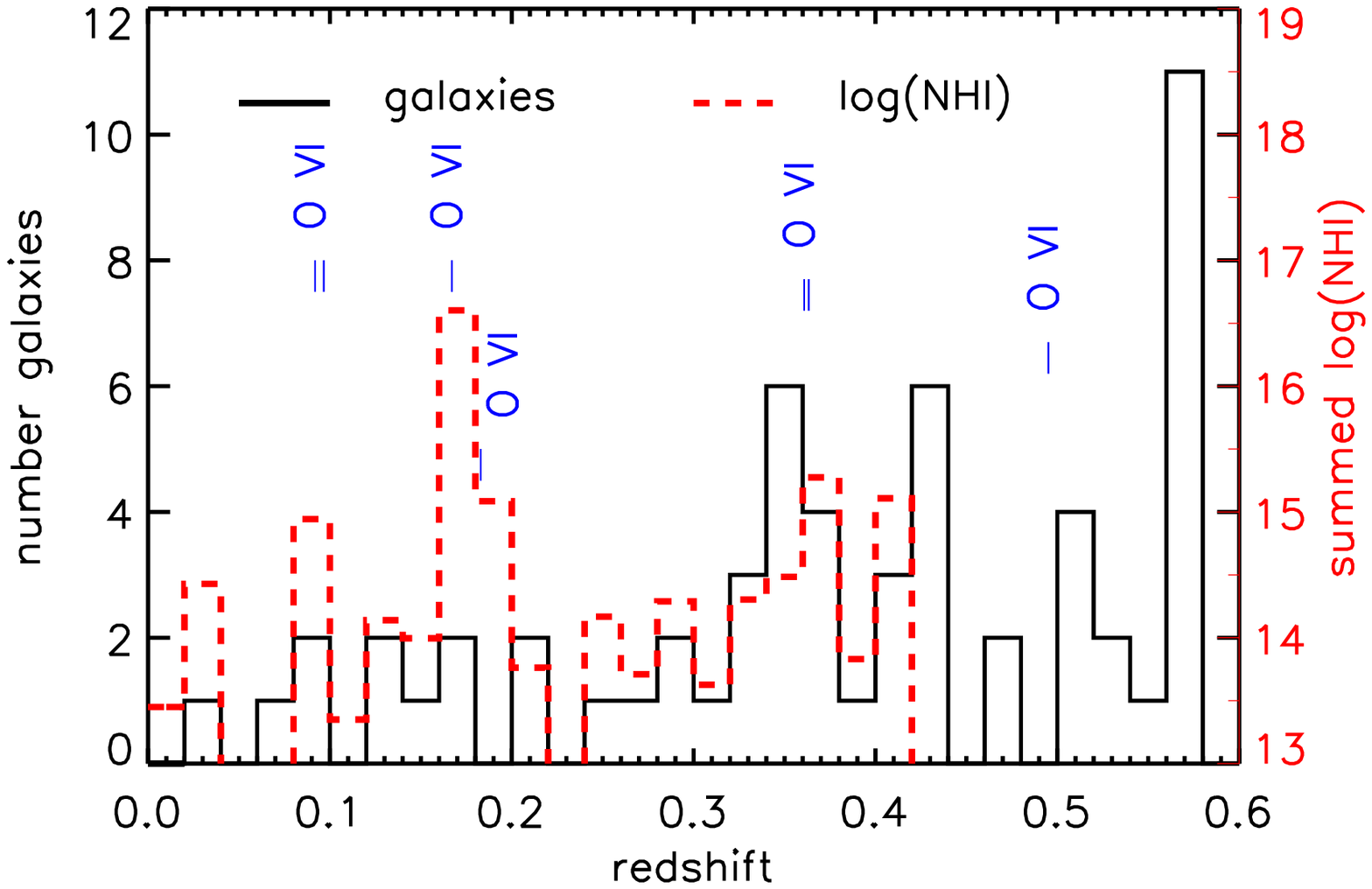} 
\caption[Galaxies {\it vs.} $\log \NHI$ toward PKS~0405--123]{
Galaxy counts (solid lines, left axis) and summed \Lya\ forest \HI\ column density
(dashed lines, right axis) counting $\log \NHI \geq 13.3$ at $0.002<z<0.423$, as a
function of redshift toward PKS~0405--123. Redshifts of \OVI\ absorbers are indicated
with ticks. Although the galaxy counts are incomplete, and the STIS \Lya\ forest data
are limited to $z<0.42$, there is a striking correlation between the local galaxy
density and local \HI\ column density in the \Lya\ forest. The Spearman rank coefficient
for this binning is $\rho=0.67$ with a two-tailed probability of $p_\rho=0.001$.  
The Kendall rank coefficient is
$\tau=0.56$ with a corresponding two-tailed probability of $p_\tau = 0.0004$.
\label{fig:nhigalhist}
}
\end{figure}

\end{document}